\documentclass[conference,compsoc]{IEEEtran}
\usepackage[numbers]{natbib}
\usepackage{amsthm}
\usepackage{amsmath}
\usepackage{subcaption}
\usepackage{amssymb}
\newcommand{\eg}{\emph{e.g.,~}}
\newcommand{\ie}{\emph{i.e.,~}}

\newcommand{\sys}{{\sc Siren}~}

\usepackage[pagebackref=true,breaklinks=true,colorlinks,bookmarks=false]{hyperref}
\usepackage{graphicx}
\usepackage{booktabs}
\usepackage{multirow}
\usepackage{makecell}
\usepackage{pifont}
\usepackage{xcolor}
\definecolor{dg}{rgb}{0,0.694,0.298}

\usepackage{breakurl}
\usepackage{bm}

\newtheorem{definition}{Definition}
\usepackage{algorithm}
\usepackage{algpseudocode}

\usepackage{enumitem}

\ifCLASSOPTIONcompsoc
  \usepackage[nocompress]{cite}
\else
  \usepackage{cite}
\fi
\ifCLASSINFOpdf
\else
\fi
\begin{document}
%
\title{Towards Reliable Verification of Unauthorized Data Usage in \\ Personalized Text-to-Image Diffusion Models}



%
\author{
    \IEEEauthorblockN{
        Boheng Li\textsuperscript{1}, 
        Yanhao Wei\textsuperscript{2}, 
        Yankai Fu\textsuperscript{2}, 
        Zhenting Wang\textsuperscript{3}, \\
        Yiming Li\textsuperscript{1}\textsuperscript{*}, 
        Jie Zhang\textsuperscript{4}\textsuperscript{*}, 
        Run Wang\textsuperscript{2}, 
        Tianwei Zhang\textsuperscript{1}
    }
    \IEEEauthorblockA{\textsuperscript{1}Nanyang Technological University, Singapore, {boheng001}@e.ntu.edu.sg, \{ym.li,tianwei.zhang\}@ntu.edu.sg }
    \IEEEauthorblockA{\textsuperscript{2}School of Cyber Science and Engineering, Wuhan University, China, \{yanhaowei,yankaifu,wangrun\}@whu.edu.cn}
    \IEEEauthorblockA{\textsuperscript{3}Rutgers University, USA, zhenting.wang@rutgers.edu}
    \IEEEauthorblockA{\textsuperscript{4}CFAR and IHPC, A*STAR, Singapore, zhang\_jie@cfar.a-star.edu.sg}
    \IEEEauthorblockA{\textsuperscript{*}Corresponding authors}
}


\maketitle

\begin{abstract}
Text-to-image diffusion models are pushing the boundaries of what generative AI can achieve in our lives. Beyond their ability to generate general images, new personalization techniques have been proposed to customize the pre-trained base models for crafting images with specific themes or styles. Such a lightweight solution, enabling AI practitioners and developers to easily build their own personalized models, also poses a new concern regarding whether the personalized models are trained from unauthorized data. 
A promising solution is to proactively enable data traceability in generative models, where data owners embed external coatings (\eg image watermarks or backdoor triggers) onto the datasets before releasing. Later the models trained over such datasets will also learn the coatings and unconsciously reproduce them in the generated mimicries, which can be extracted and used as the data usage evidence. However, we identify the existing coatings cannot be effectively learned in personalization tasks, making the corresponding verification less reliable. 

In this paper, we introduce {\sc Siren}, a novel methodology to proactively trace unauthorized data usage in black-box personalized text-to-image diffusion models. Our approach optimizes the coating in a delicate way to be recognized by the model as a feature relevant to the personalization task, thus significantly improving its learnability. We also utilize a human perceptual-aware constraint, a hypersphere classification technique, and a hypothesis-testing-guided verification method to enhance the stealthiness and detection accuracy of the coating. The effectiveness of \sys is verified through extensive experiments on a diverse set of benchmark datasets, models, and learning algorithms. \sys is also effective in various real-world scenarios and evaluated against potential countermeasures. Our code is publicly available \href{https://github.com/AntigoneRandy/SIREN}{here}.
\end{abstract}

\section{Introduction}
Modern text-to-image diffusion models \citep{rombach2022stable,runwayml2024stable,luo2023latent,razzhigaev2023kandinsky} have revolutionized the generative AI technology. Large pre-trained diffusion models, such as Stable Diffusion \citep{runwayml2024stable}, have demonstrated remarkable capabilities to produce high-quality and diverse images based on users' prompts, leading to new paradigms for commercial art and design generation.

In addition to their remarkable capabilities in generating general images, there is a growing interest in customizing these models to produce images in specific themes (e.g., generate drawings mimicking a specific art style) \citep{ruiz2023dreambooth,kumari2023multi,han2023svdiff}. This is typically achieved by fine-tuning a pre-trained model with a reference dataset. With the development of more advanced personalization techniques, the mimicry images produced by these \emph{personalized models} have become increasingly realistic and closely aligned with the desired thematic styles. Consequently, numerous real-world personalized generative AI
platforms and ecosystems \citep{civitai2024,replicate2024,scenario2024,liblib_art} have rapidly emerged, enabling personalized model trainers to share their carefully tuned personalized models or services either for free or profit. This makes the use of personalized models more accessible to a broader audience.

The remarkable success of personalized text-to-image diffusion models heavily depends on the availability of high-quality training data. However, there is a growing concern about the unauthorized usage of training data for these models \citep{wang2023diagnosis,cui2023diffusionshield}. Artists, for instance, fear that their work might be used to train these models without permission, leading to users generating images in their distinctive style and violating their copyrights \citep{shan2023glaze}. Similarly, data owners are concerned that their datasets might be exploited to train personalized models for profit, beyond the initial terms and conditions that restrict usage to specific non-commercial purposes (\eg educational) \citep{li2023black}. When a suspicious model capable of generating highly similar mimicries comes into vision, data owners may suspect unauthorized use but lack persuasive evidence to prove it, complicating efforts to formally request deletion or further pursue legal action.

One emerging solution for the aforementioned problem is to enable the \emph{traceability} of data \citep{sablayrolles2020radioactive,li2023black,guo2024domain,wang2023diagnosis,tang2023did,li2022untargeted,yu2021artificial,zhao2023recipe,luo2023steal}. The key idea is to proactively embed a special \emph{coating} (\ie secret and unique information) into the data before releasing them. This coating is imperceptible to human beings and will not interfere with visualization or other normal usage. However, it leaves a strong signal in the model trained on the coated data, which can be later detected by a specific extraction algorithm. In this paper, we explore how to enable data traceability in state-of-the-art text-to-image diffusion models. For better practicality, we consider this problem in a strict black-box setting where only generated mimicries are available (\eg through querying online APIs). Moreover, the victim/defender is assumed to have no knowledge of the infringer's training details, such as algorithms, parameters, and base models.

Prior research literature on data usage verification in ML models mainly focus on classification tasks \citep{sablayrolles2020radioactive,li2023black,guo2024domain,tang2023did,li2022untargeted,maini2020dataset,yu2021artificial,zhao2023recipe}. Only recently, researchers have tried to extend these methods to generative models \citep{yu2021artificial,zhao2023recipe,luo2023steal,wang2023diagnosis}. Some studies \citep{yu2021artificial,zhao2023recipe,luo2023steal} observe that image watermarks can be transferred from the training dataset to the output images of generative models, suggesting the potential for tracking data usage. Another line of work \citep{wang2023diagnosis} utilizes backdoor triggers to serve as the coating and trains a binary classifier to determine data ownership by detecting triggers on generated mimicries. However, these methods either are only validated to be effective in small-scale models trained from scratch \citep{yu2021artificial,zhao2023recipe}, or rely on additional assumptions about the attacker's training process \citep{wang2023diagnosis}. Unfortunately, our preliminary experiments in Section \ref{sec:motivation} reveal that these forms of coating are much harder to learn and lose effectiveness when applied to large-scale pre-trained models or when the underlying assumption is removed. These limitations lead to an important question: \emph{how to design a reliable coating that can be easily learned during personalized training?} This is particularly challenging because the learning dynamics of deep learning models are inherently complex and opaque, making it difficult for humans to analyze or even control. 

In this paper, we attempt to answer this question for the first time. Our approach, dubbed {\sc Siren}, is driven by a unified insight into the fundamental limitations of existing methods: both image watermarks and backdoor triggers focus on stealthiness while being independent of the personalized learning task. Given that large-scale pre-trained diffusion models (\eg Stable Diffusion) possess general knowledge of text and image, existing coatings are viewed as external features irrelevant to the learning task and are thus largely ignored by the model during fine-tuning. Built upon this understanding, we propose to optimize the coating to encourage the alignment between the target image and its corresponding prompt in the diffusion model feature space. In this way, the coating will carry some personalization-related features, making it more easily learned and preserved during training. However, incorporating such features usually requires larger perturbation, making the coating less imperceptible. To enhance imperceptibility and detection accuracy, we design a perceptual constraint based on the characteristics of the human visual system and jointly train a hypersphere classification-based extractor network to better extract the coating from the mimicries. By doing so, the coating remains imperceptible to human eyes but can be successfully transferred to the generated mimicries and detected by the extractor for data usage verification. We apply a hypothesis-test-guided verification technique to enhance the verification confidence. Additionally, we propose a meta-learning-based method to achieve fast adaptation to new data, making the training of \sys  more stable and efficient. 

We conduct extensive experiments on 5 state-of-the-art text-to-image diffusion models, 6 benchmark datasets, with 4 personalization learning methods. The results show that our \sys is highly effective and significantly outperforms 3 existing baselines. Specifically, \sys achieves almost 100\% true positive rates at very low significance levels across nearly all evaluated scenarios, including two real-world personalization-as-a-service platforms. It exhibits high transferability across various training algorithms, training prompts, and base models, and remains effective even when the coated data constitute only a small fraction of the entire training set. Both qualitative and quantitative evaluations, as well as a human preference study, verify that \sys has minimal impact on the visual quality of the protected images and the generation quality of the model. We also designed various potential countermeasures and validated the robustness of \sys against them.

To summarize, we make the following key contributions:
\begin{itemize}[leftmargin=*]
    \item  We take a closer look at the data usage verification problem in state-of-the-art personalized text-to-image diffusion models, and identify a shared fundamental limitation of existing solutions: the coatings are designed heuristically, without considering their relation to the learning task.
    \item  We introduce {\sc Siren}, an effective and novel methodology to trace data usage proactively in state-of-the-art personalized text-to-image diffusion models. With the help of several technical innovations, \sys significantly improves the learnability of coatings while keeping them human-imperceptible and utility harmless.
    \item We systematically validate \sys on various datasets, models, and personalization algorithms. We also show the effectiveness of \sys in various real-world scenarios, including two personalization-as-a-service platforms. We validate its robustness under several real-world scenarios as well as potential (adaptive) countermeasures.
\end{itemize}

\section{Background \& Related Work}
\noindent\textbf{Text-to-image Diffusion Models.} Recently, diffusion models have achieved remarkable advancements in image synthesis  \citep{rombach2022stable,runwayml2024stable,luo2023latent,razzhigaev2023kandinsky}. Stable Diffusion \citep{runwayml2024stable}, which is based on the latent diffusion model architecture \citep{rombach2022stable} and pre-trained on large scale text-image data, is currently the most prominent open-source text-to-image diffusion model family. This model conducts the diffusion process within a latent space generated by a pre-trained autoencoder, enabling it to leverage the highly compressed semantic features and visual patterns that the encoder has learned, thereby enhancing the efficiency and effectiveness of the image synthesis process.

\vspace{0.3em}
\noindent\textbf{Personalized Learning.} Pre-trained diffusion models, also known as base models, are good at generating generic images but are poor in customized generation needs (\eg generating specific anime characters or mimicry art style that never appeared or appeared very few times in the pre-training dataset).  To this end, both academic and industry communities are interested in fine-tuning the base model into personalized models that can generate images in specific themes or styles. Besides the standard fine-tuning method, researchers have developed advanced personalization methods \citep{ruiz2023dreambooth,kumari2023multi,han2023svdiff} to further enhance mimicking quality. 

Overall, training a decent personalized model necessitates the collection of high-quality datasets, careful adjustment of training parameters, and significant computational resources. This process can be challenging for normal users. Consequently, numerous model-sharing platforms have emerged, such as CivitAI \citep{civitai2024}, Replicate \citep{replicate2024}, and LiblibAI \citep{liblib_art}. These platforms allow model trainers to share their personalized models with others by providing either entire model weights for local reproducing, or APIs as remote services. This democratization of access to personalized models fosters a collaborative environment, allowing a wider audience to benefit from advanced AI-generated imagery.

\begin{table*}[t]
    \centering
    \caption{\small Evaluation results of backdoor-based data ownership verification method DIAGNOSIS \citep{wang2023diagnosis}.}
    \label{tab:diagnosis-lim}
    \footnotesize
    \vspace*{-0.2em}
    \setlength{\tabcolsep}{6pt}
    \resizebox{0.97\linewidth}{!}{
        \begin{tabular}{c|ccc|ccc|ccc|ccc}
            \toprule
             \multirow{3}{*}{\bf{Training Prompt}}& \multicolumn{3}{c|}{\bf{Pokemon}} & \multicolumn{3}{c|}{\bf{CelebA-HQ}} & \multicolumn{3}{c|}{\bf{ArtBench}} & \multicolumn{3}{c}{\bf{Landscape}}  \\
            \cmidrule(lr){2-4} \cmidrule(lr){5-7} \cmidrule(lr){8-10} \cmidrule(lr){11-13} 
             & DSR $\uparrow$ & DINO $\uparrow$ & FID $\downarrow$ & DSR $\uparrow$ & DINO $\uparrow$ & FID $\downarrow$ & DSR $\uparrow$ & DINO $\uparrow$ & FID $\downarrow$ &  DSR $\uparrow$ & DINO $\uparrow$ & FID $\downarrow$\\
            \midrule
            Backdoored Prompt & 100\% & 0.645 & 134.64 & 92\% & 0.551  & 84.98 & 100\% & 0.288 & 205.47 & 100\% & 0.403 & 126.76\\
            BLIP-Generated Prompt & 4\% & 0.712 & 107.42 & 4\% & 0.565 & 74.47 & 15\% & 0.294 & 209.22 & 7\% & 0.424 & 115.01\\
            \bottomrule
        \end{tabular}
    }
\end{table*}

\vspace{0.3em}
\noindent\textbf{Defending against Unauthorized Data Usage.} As personalized models start to flourish, there are growing concerns about whether these models are trained using unauthorized data \citep{cui2023diffusionshield,wang2023diagnosis,somepalli2023diffusion}. Although pre-trained base models typically open-source their pre-trained datasets \citep{stabilityai_stable_diffusion_2_1}, the rampant personalized model usually did not disclose their training data, making identifying potential infringement challenging. Existing defenses against unauthorized data usage can be broadly classified as adversarial-based and verification-based. The adversarial-based defenses \citep{van2023anti,shan2023glaze,liu2024metacloak} aim to slightly perturb the data in a way that diffusion models cannot correctly learn the desired features. For example, the state-of-the-art work Glaze \citep{shan2023glaze} adds a small, carefully designed noise onto the artwork, so that the models trained on it will learn significantly different art styles instead of the real one. Though very smart and effective, adversarial-based methods also prevent authorized training on protected data. Therefore, it mainly serves those who want to ban any model from learning from it. However, some artists or organizations may be willing to share their data for non-commercial purposes (\eg promoting academic research on generative models) but solely don't want them to be used for profit. In this scenario, they may prefer to trace the usage of data rather than rendering them totally useless for training. In contrast, the verification-based methods \citep{yu2021artificial,zhao2023recipe,wang2023diagnosis,luo2023steal} offer a more flexible solution by allowing selective detection of data usage, rather than entirely preventing models from learning from the data. One intuitive verification method is to directly detect whether the suspicious model was trained on protected data, using techniques such as membership inference \citep{pang2023black,duan2023diffusion}, or to judge whether the generated mimicries share a high style-level feature similarity using automatic models such as CLIP \citep{radford2021learning} or DINO \citep{oquab2024dinov}. However, it remains challenging to obtain satisfactory performances, due to the inherent complexity and generalizability of AIGC models (see a detailed discussion in Appendix \ref{app:discussion}). As such, researchers also focus on proactive solutions \citep{luo2023steal,wang2023diagnosis}, which rely on external coatings (\ie image watermarks or backdoor triggers) to trace data usage. However, current proactive verification methods are limited to small-scale models or rely on additional assumptions, which we will identify to be not enough for this important yet challenging problem.

\begin{figure}[t]
    \centering
    \begin{subfigure}{0.235\textwidth}
        \centering
        \includegraphics[width=\linewidth]{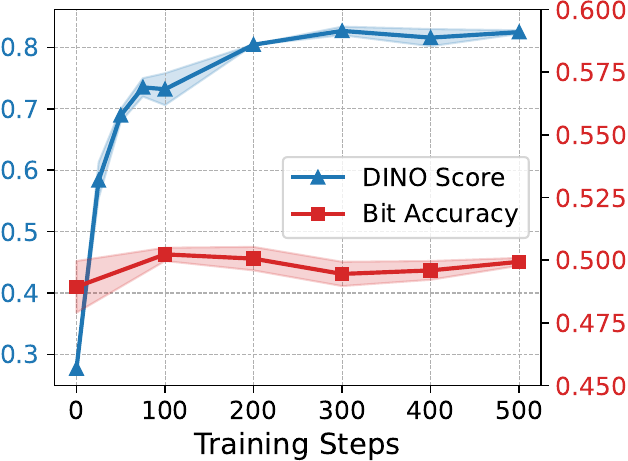}
        \caption{\citet{yu2021artificial}}
        \label{fig:yu}
    \end{subfigure}
    \begin{subfigure}{0.235\textwidth}
        \centering
        \includegraphics[width=\linewidth]{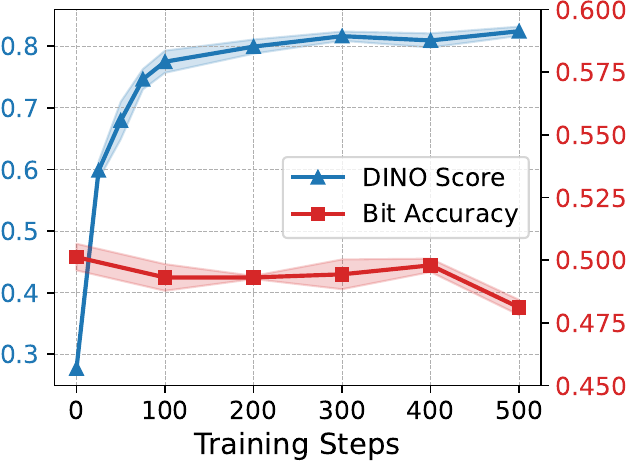}
        \caption{\citet{luo2023steal}}
        \label{fig:luo}
    \end{subfigure}
    \caption{\small Evaluation results of watermark-based methods on Dog \citep{ruiz2023dreambooth} dataset. The personalization method is DreamBooth \citep{ruiz2023dreambooth}. The model quickly learns the new concept while ignores the watermark. The bit accuracy would be $50\%$ for random guesses.}
    \label{fig:luoandyu}     
    \vspace{-1.5em}
\end{figure}

\section{Motivating Studies}
\label{sec:motivation}
\subsection{Watermark-based Methods}
\label{sec:watermark-lim}

Existing watermark-based methods \citep{yu2021artificial,zhao2023recipe,luo2023steal} embed a pre-defined steganography message into the protected images, with the expectation that the same message can be decoded from images generated by the personalized model trained on them. While previous research has demonstrated the feasibility of these solutions in GANs \citep{yu2021artificial} and small-scale diffusion models (\eg DDPM) trained from scratch \citep{zhao2023recipe}, as we will verify, they become less effective in the context of personalized learning with state-of-the-art Stable Diffusion model. Below we conduct experiments to show that image watermarks cannot be adequately learned during the fine-tuning process, even though the personalized models are already capable of producing high-quality mimicries. 

We reproduce and evaluate the watermark-based methods of \citet{yu2021artificial} and \citet{luo2023steal} in this section. Specifically, we fine-tune the Stable Diffusion v1.5 model \citep{rombach2022stable} using the DreamBooth method \citep{ruiz2023dreambooth} on a benchmark datasets for personalization learning: Dog \citep{ruiz2023dreambooth}. Following \citep{yu2021artificial,luo2023steal}, we watermark all images in the training set with the same pre-defined bit message and then use them for personalized learning. We then generate 1,000 mimicry images and extract the watermark from them. The effectiveness of watermarks is evaluated using the {Bit Accuracy} metric, defined as the ratio of successfully extracted bits to the total number of bits.  Additionally, we assess the quality of personalization learning using the {DINO score} \citep{caron2021emerging}, which is the average pairwise cosine similarity between the ViTS/16 DINO embeddings of generated and real images. It is widely used to evaluate the effectiveness of personalized learning \citep{ruiz2023dreambooth}. A higher DINO score indicates greater semantic similarity and, therefore, better mimicry performance.

We repeat each experiment 3 times and report the results in Figure \ref{fig:luoandyu}. As shown, the model quickly learns the new concept during personalized learning and starts to produce high-quality mimicry images within 100 to 200 time steps, indicated by a quick increase in DINO score. However, the watermarks are largely ignored during personalization learning. Even at the 500th time step, where we observe the model starts to slightly overfit the training set, the bit accuracy of both watermarks remains around $50\%$. This indicates that the watermarks fail to be preserved.

To summarize, existing watermarks \citep{yu2021artificial,luo2023steal} are difficult to be learned and preserved during personalization learning with state-of-the-art Stable Diffusion models, \ie they have only limited \emph{learnability}. We hypothesize that one fundamental reason is these models have been pre-trained on large high-quality text-image datasets, leading to the establishment of robust semantic connections between text and image concepts. In other words, they have been familiar with general concepts and potentially know \emph{what to learn} when presented with new concepts. For example, pre-trained diffusion models are already familiar with the general appearance of dogs. When adapting to a specific type of new dog, the model may mainly focus on the distinctive details of such new dog, such as fur, eyes, and ears. However, image watermarks have limited semantic connections to the primary concepts. As a result, they will possibly be treated as extraneous features similar to image backgrounds by the model and consequently disregarded during training.

\subsection{Backdoor-based Methods}

DIAGNOSIS \citep{wang2023diagnosis} is currently the only  backdoor-based data usage tracing method effective in text-to-image diffusion models. It coats the dataset by adding a stealthy backdoor trigger  onto protected images, and appending a  trigger text (\ie a rarely used word, such as ``tq'') to the corresponding original prompt. Then, if a model is trained on this coated dataset, it will learn a ``backdoor'' (\ie to add the same backdoor trigger on the generated images if the trigger text is met). By training a binary classifier on external datasets and using it to detect whether the mimicries contain the backdoor trigger, the defender can determine whether the suspicious model was trained on protected data. 

One underlying assumption of DIAGNOSIS is that \emph{the training prompt used by the infringer must be the backdoored one provided by the defender}. However, this assumption can be easily bypassed without harming the generation quality -- the infringer can use state-of-the-art image captioning models, such as BLIP \citep{li2023blip}, to generate high-quality, detailed text descriptions as training prompts. Unfortunately, we will show that when this assumption is removed, the effectiveness of DIAGNOSIS will reduce significantly.

We reproduce DIAGNOSIS on four benchmark datasets, \ie Pokemon \citep{pratama2019pokemon}, CelebA-HQ \citep{karras2018progressive}, ArtBench \citep{liao2022artbench}, and Landscape \citep{arnaud58_landscape_pictures}. Specifically, we fine-tune the Stable Diffusion v1.5 model \citep{runwayml2024stable} under two settings: {(a)} both images and prompts are coated (Backdoored Prompt);  and {(b)} only images are coated, but the prompts are generated using the BLIP image captioning model \citep{li2023blip} (BLIP-Generated Prompt);  To measure whether DIAGNOSIS is successful, we use 3 metrics: Detection Success Rate (DSR) which is the ratio of mimicries that are correctly classified  as ``contains the trigger'' to measure effectiveness, along with DINO score \citep{caron2021emerging} and FID \citep{heusel2017gans} to evaluate the generation quality.

As can be seen in Table \ref{tab:diagnosis-lim}, when the infringer uses the BLIP-generated prompts, the quality of the mimicries remains comparable to that trained with backdoored prompts. However, the DSRs drop quickly on all datasets. We also validate DIAGNOSIS using Welch's T-test \citep{welch1947generalization}, and the results confirm that the difference in DSR is not statistically significant compared to an independent clean model. This suggests that the success of DIAGNOSIS is heavily dependent on the assumption that the infringer uses exactly the same training prompts provided by the defender. The original DIAGNOSIS paper mitigates this issue by sacrificing both training set quality and generation quality: it enlarges the trigger strength to twice that of the original so that the trigger becomes visible and will be preserved even when the infringer does not use the backdoored prompt. However, as we will validate in our experiments, this remedy is only effective on certain datasets and personalization methods. 

In conclusion, DIAGNOSIS encounters a similar \emph{learnability issue} when the assumption about training prompts is removed. We believe the underlying reason is similar to our analysis in Section \ref{sec:watermark-lim}: backdoor triggers are designed heuristically, without considering their correlation to the personalization task. When the training prompts include the text trigger, the model can correctly associate the image triggers with it. However, when such text triggers are not contained, the model barely considers backdoor trigger as a feature relevant to the personalization task. As a result, the triggers are also largely ignored during training.

\section{\sys}
\subsection{Threat Model}
We consider a practical scenario involving three parties: a data owner (victim), an infringer (attacker), and a third-party data protection platform (defender). 

\vspace{0.3em}
\noindent\textbf{Data Owner's Capabilities \& Goals.} The data owner aims to release his/her possessed images to the public for certain purposes (\eg artwork advertising or promoting academic research). However, he/she does not want his/her data to be used for commercial purposes without authorization, \ie training and selling personalized diffusion models for profit in our consideration. To protect the data, the user can request a third-party platform to coat the images before releasing them. When the data owner observes a black-box suspicious model, they can ask the platform to verify any potential infringements of such models. 

\vspace{0.3em}
\noindent\textbf{Infringer's Capabilities \& Goals.} The infringer (also the attacker) aims to develop a personalized diffusion model capable of generating high-quality mimicry images. To this end, he/she needs to collect some data from the Internet following his/her desired concept or style. He/She obtains the data owner's protected (\ie coated) images and uses them as (part of) the training dataset. We assume the attacker (1) has complete access and control to the collected dataset, (2) has complete control over the fine-tuning and generation procedure, and (3) has knowledge that the collected data is (possibly) coated. However, he/she (4) needs to ensure that the generated mimicries are with high-quality, and (5) may know the design of \sys (in an adaptive attack setting) but cannot access the exact network parameters of the coating generator and extractor used by the defender.

\vspace{0.3em}
\noindent\textbf{Data Protection Platform's Capabilities \& Goals.} This platform is a trusted third party, providing registered users with data coating and verification services. The platform has (1) complete access and control over the data provided by the owner, so it can add special coatings onto the data before releasing it; and (2) black-box access to the suspicious model, so it can query the suspicious model and obtain the generated mimicries for verification. However, the platform (3) needs to keep the coated data visual and utility similar to the uncoated version, and (4) does not know or control any training details (\eg base model, training prompts, personalization methods) of the suspicious model.

\subsection{Design Overview}
Similar to previous methods, the framework of \sys is divided into two stages: coating and verification. During the coating stage, the defender jointly trains a coating generator $\mathcal{G}: \mathbb{R}^{c\times h\times w}\rightarrow\mathbb{R}^{c\times h\times w}$, which takes an image as input and produces a coating of identical size, and a paired coating extractor $\Phi: \mathbb{R}^{c\times h\times w}\rightarrow\mathbb{R}$ to detect the coating from a suspicious image and output a specific coating score (explained later). Once training is complete, the defender generates a unique coating for each image in the dataset, applies these coatings, and returns the coated dataset back to the user. In the verification stage, when a suspicious black-box personalized model is observed, the user can request the protection platform to verify whether this model incorporates the coated images for personalization training. The platform queries the suspicious model to obtain the generated mimicries, calculate the coating score, and conduct a hypothesis test to make the decision.

We make several innovations in the design of \sys to enhance its practicality and effectiveness. (1) To enhance learnability, we design a novel learnability loss by correlating the coating to the personalized learning process (Section \ref{sec:learnability-loss}). (2) To enhance the stealthiness of the coated images, we introduce the HVS-aware perceptual constraint, which leverages the Human Visual System to reduce the visual distortions (Section \ref{sec:hvs-aware-constraint}). (3) We introduce the hypersphere classification loss (Section \ref{sec:hypersphere-classification}) and distributional hypothesis testing (Section \ref{sec:hypothesis-testing}) to detect the usage of coated data. (4) We further propose a meta-learning technique to boost the training of the coating generator and extractor (Section \ref{sec:meta-learn}). Below we give the design details of each technique. 

\subsection{Training \& Coating Stage}
In the training stage, the defender jointly trains a coating generator $\mathcal{G}$ and a paired extractor ${\Phi}$. Below we first introduce several essential loss terms used in this stage and present the overall training objective.

\vspace{-1em}
\subsubsection{Learnability Loss}
\label{sec:learnability-loss}
Motivated by the limited learnability of existing coatings, the key intuition behind our solution is to ensure that \textit{the traceable coating is relevant to the features that personalized learning wants to learn}.  In other words, we want the coating itself to be a relevant feature that is helpful for personalization and can be effectively learned by the diffusion model during fine-tuning to reproduce it in the mimicries. Although intuitively reasonable, achieving this goal is challenging in practice since the learning dynamics of large diffusion models are complex and even difficult to analyze, not to mention controlling them. 

To this end, we formulate an optimization problem to obtain the desired coating. Before stepping into the details, we first conceive a definition of feature-relevant coating.
\vspace{-0.3em}
\begin{definition}[Feature-relevant Coating]
For a personalized model $\epsilon_{\theta}^*$, a training image-text pair $({x},{t})$, and $\tau > 0$, a coating $\delta$ is $\tau$-feature-relevant if:
\begin{equation}
\mathcal{L}_{\text{DM}}({x},{c})-\mathcal{L}_{\text{DM}}({x}+\delta,{c}) =\tau,
\end{equation}
where $\mathcal{L}_{\text{DM}}(\cdot,\cdot)$ is the loss function of the target diffusion model (DM). For latent diffusion models, the loss function is $\mathcal{L}_{\text{LDM}}=\mathbb{E}_{\epsilon \sim \mathcal{N}(0,1), t} \left[ \| \epsilon - \epsilon^*_{\theta}(z_t, t, c) \|_2^2 \right]$, where $\epsilon^*_{\theta}(\cdot,\cdot,\cdot)$ is the target diffusion model, $z_t$ is the noised latent representation of the image and $t$ is the time step \citep{rombach2022stable}.
\end{definition}
Similar to previous studies on feature-learning theory and adversarial attacks \citep{ilyas2019adversarial,tsipras2018robustness}, this definition states that a coating is relevant to the features of a training image if patching it to the training sample can reduce the loss of this text-image pair on the model. For an intuitive understanding, pre-trained text-to-image diffusion models have already established a robust, text-image aligned feature space \citep{kwon2022diffusion,li2023your}. In this context, the loss of a given text-image pair represents the semantic discrepancy between the text and image considered by the model. If adding the coating to the image reduces this loss, it implies that the coating encourages the alignment between the text and the image. For instance, if a coating reduces the loss between an image of a dog and the prompt `dog', it means that the coating contains some features recognized by the diffusion model as the characteristic of a dog.

Intuitively, a coating with larger $\tau$ indicates higher relevance to the target feature. Therefore, our goal is to identify a coating with the largest possible $\tau$. Thus, given the target dataset $\mathcal{D}=\{(x_i,c_i)\}_{i=1}^{N}$, we aim to minimize the learnability loss, defined as:
\begin{equation}
\begin{aligned}
\label{eq:learn}
{\mathcal{L}_{\text{learn}}}&=-\frac{1}{N}\sum_{(x_i,c_i)}{(\mathcal{L}_{\text{DM}}({x_i},{c_i})-\mathcal{L}_{\text{DM}}({x_i+\mathcal{G}(x_i)},{c_i}))},
\end{aligned}
\end{equation}
where $\mathcal{G}(\cdot)$ is the coating generator. Note that the above optimization problem requires white-box access to the personalized model and the corresponding ground-truth prompt used by the infringer, which is often not realistic in the real world. To relax this requirement, inspired by previous works \citep{shan2023glaze}, we employ a surrogate diffusion model to approximate $\epsilon_{\theta}^*$. In detail, we fine-tune the Stable Diffusion v1.5 \citep{rombach2022stable} on the uncoated dataset for a few (30 in this paper) epochs and use this model to serve as $\epsilon_{\theta}^*$.  Additionally, we follow the approach in \citep{ruiz2023dreambooth} to derive the class descriptor as the surrogate prompt for both fine-tuning the surrogate model and calculating Eq. (\ref{eq:learn}). Our experiments demonstrate that despite the use of surrogates, \sys exhibits high transferability across various diffusion models (including those with completely different architectures) and diverse training prompts.

\begin{figure}[t]
    \centering
    \begin{subfigure}{0.15\textwidth}
        \centering
        \includegraphics[width=\linewidth]{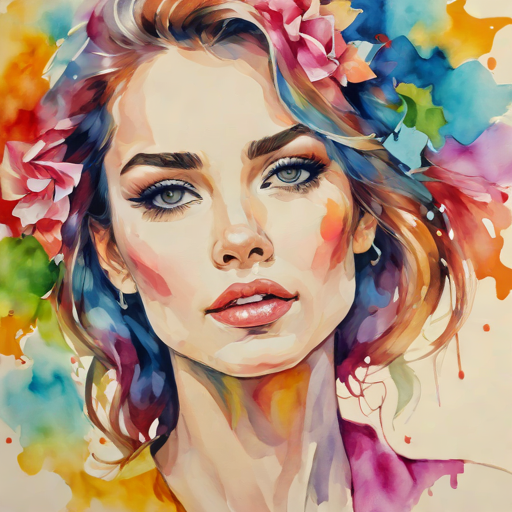}
        \caption{Original}
        \label{fig:original}
    \end{subfigure}
    \hfill
    \begin{subfigure}{0.15\textwidth}
        \centering
        \includegraphics[width=\linewidth]{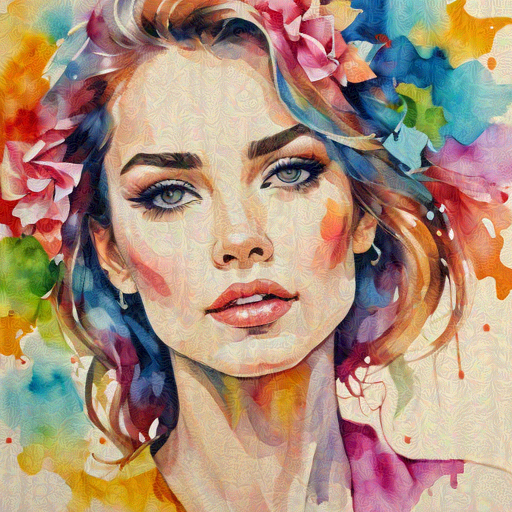}
        \caption{$\ell_{\infty}$ Constraint}
        \label{fig:linf}
    \end{subfigure}
    \hfill
    \begin{subfigure}{0.15\textwidth}
        \centering
        \includegraphics[width=\linewidth]{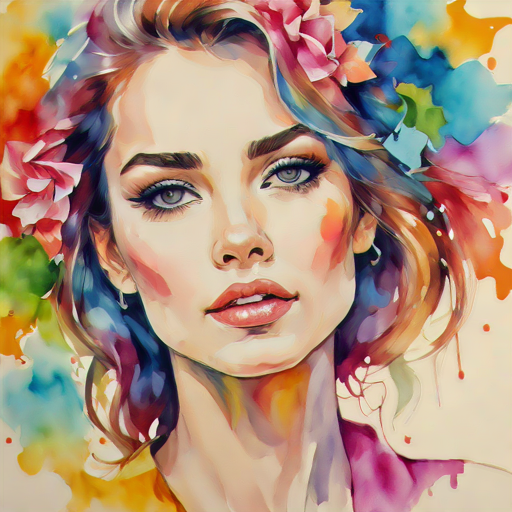}
        \caption{Ours}
        \label{fig:color}
    \end{subfigure}
    \vspace{-0.5em}
    \caption{\small A comparison of (a) original uncoated image; with images coated by (b) the $\ell_\infty$ constraint of $11/255$, and (c) our {\sc Siren}. Notably, the coating optimized by $\ell_\infty$ constraint brings unnatural artifacts on flat and bright color areas (\eg the face of the woman), while our coating looks much more natural.}
    \label{fig:colorvslinf} 
    \vspace{-1.5em}
\end{figure}

\vspace{-1em}
\subsubsection{HVS-aware Perceptual Constraint}
\label{sec:hvs-aware-constraint}
Recall that an ideal coating should be not only learnable but also stealthy. This means that when the coating is applied to the protected image, it cannot cause noticeable changes or appear unnatural to human observers. Previous works on data protection commonly constrains the coating budget to a certain $\ell_{p}$ norm \citep{van2023anti,liu2024metacloak} (\eg the state-of-the-art work \citep{liu2024metacloak} uses a budget of $\ell_{\infty} = 11/255$). However, this constraint, measured in the RGB color space, does not fully exploit the characteristics of the Human Visual System (HVS) and may cause unnatural color distortions on the coated image (Figure \ref{fig:linf}). 

Inspired by existing works on HVS \citep{luo2001development}, we employ a HVS-aware perceptual constraint, \ie the perceptual color distance, to improve the stealthiness of coatings in {\sc Siren}. This distance is quantified using the CIEDE2000 color difference formula \citep{sharma2005ciede2000}, which provides a more accurate measure of the perceived difference between images as experienced by human observers. Given two images, the perceptual color difference $\Delta E(\cdot,\cdot)$ is calculated as:
\begin{equation}
\begin{aligned}
\label{eq:deltae}
\Delta E= \sqrt{(\frac{\Delta C'}{k_C S_C})^2 +(\frac{\Delta L'}{k_L S_L})^2 +  (\frac{\Delta H'}{k_H S_H})^2 + \Delta R},
\end{aligned}
\end{equation}
where ${\Delta C'}$, ${\Delta L'}$ and ${\Delta H'}$ denote the chroma, lightness, and hue distance between two images in the CIELCH space, respectively, and $\Delta R = R_T (\frac{\Delta C'}{k_C S_C}) (\frac{\Delta H'}{k_H S_H})$ is an interactive term between chroma and hue differences. The weighting functions $S_L$, $S_C$ , $S_H$ and $R_T$ as well as other parameters $k_C$, $k_L$ and $k_H$ are derived based on large-scale human experiments to better approximate HVS perception. We refer readers to \citep{luo2001development} for more details on how the formulation is derived and why it can better simulate human perception.
We incorporate this perceptual constraint as an additional regularizer to encourage smoother color changes:
\begin{equation}
\begin{aligned}
\label{eq:colorloss}
\mathcal{L}_\text{percept}=\frac{1}{N}\sum_{i=1}^N\|\Delta E(x_i, x_i+\mathcal{G}(x_i))\|_2^2.
\end{aligned}
\end{equation}

Figure \ref{fig:colorvslinf} compares images coated by our \sys with that coated with $\ell_\infty$ constraint. The image coated with the $\ell_\infty$ constraint exhibits noticeable distortions and unnatural textures. In contrast, the image coated by \sys appears much more natural and stealthy, suggesting a notable improvement in visual quality compared with $\ell_\infty$ constraints. 

\vspace{-1em}
\subsubsection{Hypersphere Classification Loss}
\label{sec:hypersphere-classification}
By far, we have designed techniques to make the coatings stealthy to minimally impact the image's visual quality, and learnable by the diffusion models. The next goal is to detect the existence of the coating on the mimicries. A straightforward solution is to directly train a binary classifier on coated and clean images using the standard cross-entropy loss, which essentially learns a hyperplane in the classifier feature space to distinguish between positive (coated) and negative (clean) samples. However, this is sub-optimal because it is impractical to collect all possible clean images in the real world. Consequently, the learned hyperplane might be biased towards the training dataset, and may cause misclassification on unseen negative data  (see a detailed discussion in Appendix \ref{app:discussion}), as illustrated in Figure \ref{fig:oneclass} (left).

\begin{figure}[t]
    \centering
        \includegraphics[width=\linewidth]{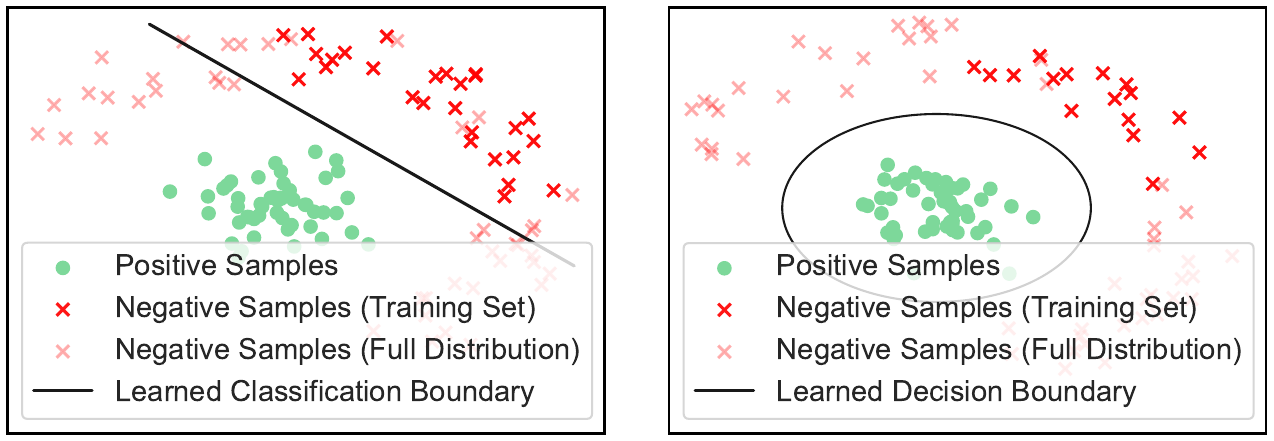}
    \vspace{-1em}
    \caption{\small An example feature-space illustration comparing binary classification (left) with hypersphere classification (right). Direct binary classification might be biased by the incomplete negative training data, while hypersphere classification mainly focuses on positive samples and generalizes better on unseen negative data.}
    \label{fig:oneclass}
    \vspace{-1.5em}
\end{figure}

Inspired by previous works on data description \citep{ruff2018deep}, we propose to learn a hypersphere rather than a hyperplane. This approach learns a minimal hypersphere that can encompass all positive samples and regard all other samples out of the hypersphere as negative. As a result, the learned boundary will mainly focus on positive samples (\ie coated images) and be much less affected by the distribution of negative training set, providing better generalizability to unseen negative samples (Figure \ref{fig:oneclass} right). Specifically, given a feature extractor $\Phi:\mathbb{R}^{c\times h \times w}\rightarrow \mathbb{R}^{d}$, the hypersphere classification loss for positive samples is:
\begin{equation}
\begin{aligned}
     \mathcal{L}_\text{hc}^{+}=\nu R^2 + \frac{1}{N} \sum_{i=1}^{N}\max\{0, \left \| \Phi({x}_i+\mathcal{G}(x_i)) -{o} \right \|_2^2 - R^2\},
\end{aligned}
\label{eqn:hcl}
\end{equation}
where $o\in\mathbb{R}^d$ and $R\in\mathbb{R}$ are the center and radius of the hypersphere, respectively, and $\nu$ is a hyperparameter controlling the relative strength of the two terms. Intuitively, $\mathcal{L}_{hc}^+$ consists of two terms: the first term minimizes the volume of the hypersphere and the second term penalizes the positive samples that are out of the hypersphere. To better leverage the negative samples, we also minimize the following objective that pushes them out of the hypersphere:
\begin{equation}
\begin{aligned}
     \mathcal{L}_\text{hc}^{-}=-\frac{1}{N} \sum_{i=1}^{N}\log(1-\exp(-\|\Phi({x}_i)-o\|_2^2)).
\end{aligned}
\label{eqn:ncl}
\end{equation}
$o$ is initialized and updated as the mean representation of all positive samples in the batch after each iteration, while $R$ is updated via line search \citep{forsythe1977computer}.

\vspace{-1em}
\subsubsection{Overall Training Objective}

Given the aforementioned losses, our overall training objective is defined as:
\begin{equation}
\label{eq:overall}
    \min_{\mathcal{G}, \Phi} \mathcal{L}_{\text{overall}}=\mathcal{L}_{\text{learn}} + \lambda_1 \mathcal{L}_{\text{percept}} + \lambda_2 (\mathcal{L}_\text{hc}^{+} + \mathcal{L}_\text{hc}^{-}),
\end{equation}
where $\lambda_1$ and $\lambda_2$ are weighting parameters that control the relative strengths of the losses. During training, we follow previous work \citep{fernandez2023stable} to include a differentiable EoT layer and an MSE image loss to enhance robustness and stabilize training. The full training algorithm can be found in Algorithm \ref{alg:sys} and more details are in Appendix \ref{app:details}.

\begin{algorithm}[t]
\small
\caption{Protecting data with \sys}\label{alg:sys}
\begin{algorithmic}[1]
\Statex \textbf{Input}: Uncoated data $\mathcal{D}$ with $N$ samples $x_1, \cdots, x_N$, meta coating generator $\mathcal{G}^*$ and extractor $\Phi^*$, learning rate $\alpha$, $\beta$
\State $\mathcal{G},{\Phi} = \text{Clone}(\mathcal{G}^*,{\Phi}^*)$
\State $o \leftarrow \frac{1}{N} \sum_{i=1}^{N}(\Phi(x_i+\mathcal{G}(x_i)))$ \Comment{Initialize ${o}$}
\State $R \leftarrow \frac{1}{N} \sum_{i=1}^{N}(\|\Phi(x_i+\mathcal{G}(x_i))-o\|_2^2)$ \Comment{Initialize ${R}$}
\While{loss not converged}
        \State Sample (a batch) of $x_i$ from $\mathcal{D}$
        \State $c_i \leftarrow \text{get\_class\_descriptor}(x_i)$
        \State Calculate $\mathcal{L}_{\text{overall}}$ on $(x_i,c_i)$ with $\mathcal{G}$ and ${\Phi}$ via Eq. (\ref{eq:overall})
        \State    $\mathcal{G} \leftarrow \mathcal{G}-\alpha \nabla_{\mathcal{G}} \mathcal{L}_{\text{overall}}$
        \State ${\Phi} \leftarrow {\Phi}-\beta \nabla_{{\Phi}} \mathcal{L}_{\text{overall}}$
    \State $o \leftarrow \frac{1}{N} \sum_{i=1}^{N}(\Phi(x_i+\mathcal{G}(x_i)))$
    \State Update $R$ via line search
\EndWhile
\Statex \Return $\{x_i+\mathcal{G}(x_i)\}_{i=1}^N$
\end{algorithmic}
\end{algorithm}

\subsection{Verification Stage}
\label{sec:hypothesis-testing}

Given a mimicry image $x_s$ generated by the suspicious model, we can determine whether it contains the coating by projecting it into the feature space of $\Phi$ and calculating its distance to the center of the hypersphere, \ie $s(x_s)=\|\Phi(x_s)-o\|_2^2$, which we call the \emph{coating score}. Ideally, coated images will have small coating scores while clean ones will have much larger scores. To convert the coating scores to human-readable evidence, we follow a previous work \citep{wang2023diagnosis} and conduct a distributional hypothesis test. In detail, we have the null hypothesis $H_0$: unauthorized data usage is not detected, and the alternative hypothesis $H_1$: unauthorized data usage is detected. Given that mimicries generated by personalized models from coated data have statistically different coating scores from the clean data, we conduct a two-sample Kolmogorov–Smirnov (K-S) test \citep{massey1951kolmogorov} to determine whether the suspicious model is personalized using coated images. Given a significance level $\alpha$, we reject the null hypothesis and claim the detection of unauthorized data usage if the following inequality is satisfied:
\begin{equation}
\begin{aligned}
\label{eq:kstest}
\sup_x \left| F(x) - G(x) \right| -  \sqrt{\frac{n+m}{nm}} K_\alpha> 0,
\end{aligned}
\end{equation}
where $F(x)$ and $G(x)$ represent the empirical distribution functions of the coating scores of coated and clean samples, respectively, $n$ and $m$ are the sizes of these two samples, and $K_\alpha=2\sum_{k=1}^{\infty}(-1)^{k-1}e^{-2k^2x^2}$ is the critical value from the K-S distribution corresponding to $\alpha$ \citep{marsaglia2003evaluating}. Note that the significance level $\alpha$ models the probability of making Type-I error, namely rejecting the null hypothesis while it is actually true \emph{i.e.} the false positives. 

The benefits of  distributional hypothesis testing are as follows. First, the K-S test is a non-parametric test, which does not rely on any additional assumptions on the two distributions. Second, by setting $\alpha$ to a small value (\eg $10^{-6}$), we can reduce the FPR to enhance the credibility of \sys and prevent potential misaccusation on benign models. We verify the controlled FPR of the test and its sensitivity to different benign distributions in Appendix \ref{app:fpr}.

\begin{figure*}[t]
    \centering
    \begin{subfigure}[b]{0.24\textwidth}
        \centering
        \includegraphics[width=\textwidth]{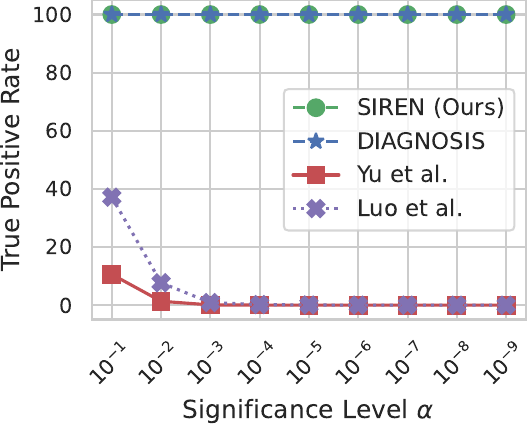}
        \caption{Pokemon \citep{pratama2019pokemon}}
    \end{subfigure}
    \hfill
    \begin{subfigure}[b]{0.24\textwidth}
        \centering
        \includegraphics[width=\textwidth]{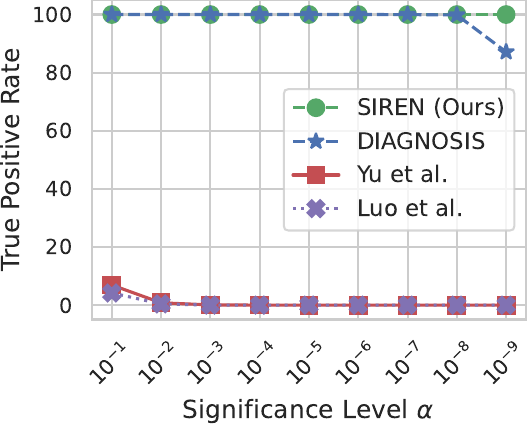}
        \caption{CelebA-HQ \citep{karras2018progressive}}
    \end{subfigure}
    \hfill
    \begin{subfigure}[b]{0.24\textwidth}
        \centering
        \includegraphics[width=\textwidth]{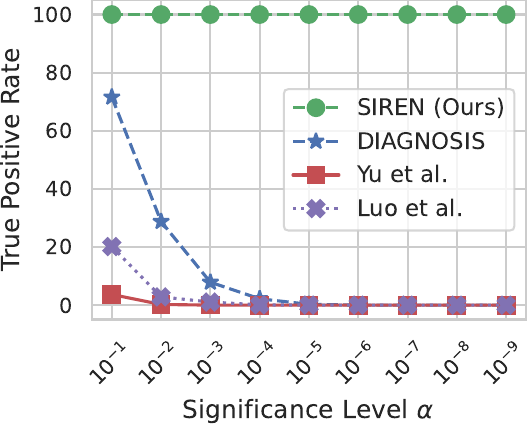}
        \caption{ArtBench \citep{liao2022artbench}}
    \end{subfigure}
    \hfill
    \begin{subfigure}[b]{0.24\textwidth}
        \centering
        \includegraphics[width=\textwidth]{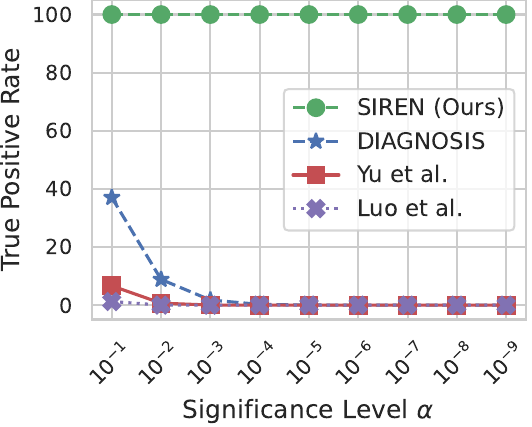}
        \caption{Landscape \citep{arnaud58_landscape_pictures}}
    \end{subfigure}
    \caption{Effectiveness comparison in the fine-tuning personalization scenarios.}
    \label{fig:ft-main}
    \vspace{-1.5em}
\end{figure*}

\subsection{Meta-learning for Fast Adaptation}
\label{sec:meta-learn}
While achieving state-of-the-art performance, the coating stage of \sys can be burdensome since we need to retrain a coating generator and coating feature extractor from scratch every time, which is time- and resource-consuming. To this end, in this section, we propose to leverage meta-learning to mitigate the aforementioned challenges. Specifically, we aim to learn a ``meta'' coating generator $\mathcal{G}^*$ and extractor $\Phi^*$ whose feature spaces are well-structured and thus their initialization weights can be easily adapted to the new coatings using a few fine-tune steps. To this end, we use a first-order meta-learning method, Reptile \citep{nichol2018reptile}. Given a batch of proxy data $\mathcal{D}_p$, we first set $\mathcal{G}^*_0=\mathcal{G}^*$ and $\Phi^*_0=\Phi^*$. Then, we fine-tune this model pair to yield an updated model pair $\mathcal{G}^{*}_K$ and $\Phi^*_K$ using $K$ steps of SGD update:
\begin{equation}
\begin{aligned}
    \{\mathcal{G}^*_k,\Phi^*_k\} = \text{SGD}(\{\mathcal{G}^*_{k-1},\Phi^*_{k-1}\}, \mathcal{D}_p), k=1,\dots, K,
\end{aligned}
\end{equation}
After this, the parameter difference of this personalized update is used as the meta-gradient to train the meta model:
\begin{equation}
\begin{aligned}
    \mathcal{G}^* &\leftarrow \mathcal{G}^* - \gamma (\mathcal{G}^*-\mathcal{G}^*_K),\\
    \Phi^* &\leftarrow \Phi^* - \xi (\Phi^*-\Phi^*_K),
\end{aligned}
\end{equation}
where $\gamma$ and $\xi$ are the meta learning rates. With this, we can gradually learn a meta model that can easily and quickly adapt to new data with very few training steps. We provide the full training algorithm and more details in Appendix \ref{app:details}.

\section{Evaluation}

\subsection{Experimental Setup}
\label{sec:expsetup}
\noindent\textbf{Diffusion Models.} We use 5 state-of-the-art text-to-image diffusion models (\ie Stable Diffusion v1.5 \citep{runwayml2024stable}, Stable Diffusion v2.1 \citep{stabilityai_stable_diffusion_2_1}, Kandinsky 2.2 \citep{razzhigaev2023kandinsky}, Latent Consistency Models \citep{luo2023latent} and VQ Diffusion \citep{gu2022vector}) in our experiments. It is worth noting that except for Stable Diffusion v2.1 which shares the same network architecture with our surrogate model but trained under different settings and datasets, the remaining models have totally different network structures (and also model parameters) compared to the surrogate. 

\vspace{0.3em}
\noindent\textbf{Personalization Methods and Datasets.} We evaluate the generalizability of \sys under 4 personalzation methods, including the fine-tuning (on 4 large scale dataset, \ie Pokemon \citep{pratama2019pokemon}, CelebA-HQ \citep{karras2018progressive}, ArtBench \citep{liao2022artbench}, and Landscape \citep{arnaud58_landscape_pictures}) and 3 advanced methods (\ie DreamBooth \citep{ruiz2023dreambooth}, SVDiff \citep{han2023svdiff}, and Custom Diffusion \citep{kumari2023multi}), on 2 relatively small datasets (\ie Dog \citep{ruiz2023dreambooth} and WikiArt subset \citep{saleh2015large}). More details on the methods and datasets can be found in Appendix \ref{app:details}. We individually protect the dataset with \sys in each setting and train the personalized model, then generate the mimicries for verification. We use LoRA \citep{hu2022lora} to save memory usage in the fine-tuning experiments. 

\vspace{0.3em}
\noindent\textbf{Evaluation Metrics.} The effectiveness of our method and baselines are assessed using the True Positive Rate (TPR) metric at certain significance level $\alpha$. This metric quantifies the proportion of correctly identified true positives at a specified significance level. For instance, if a method achieves a TPR of 97\% at  $\alpha=10^{-6}$, then it can correctly identify 97 out of every 100 really positive instances under $\alpha=10^{-6}$. Note that higher TPR at lower significance level indicates better reliability. Moreover, we use three metrics widely used in image quality assessment, namely PSNR \citep{hore2010image}, SSIM \citep{wang2004image} and LPIPS \citep{zhang2018unreasonable}, to quantitatively measure the impact of \sys on image quality. Finally, we use the CLIP score \citep{radford2021learning}, DINO score \citep{oquab2024dinov}, and FID \citep{heusel2017gans}, which are widely used by previous works \citep{ruiz2023dreambooth,han2023svdiff}, to measure the generation quality. 

\vspace{0.3em}
\noindent\textbf{Baselines.} We mainly compare our \sys with 3 state-of-the-art verification-based methods, \ie two watermarking-based (\citet{yu2021artificial}, \citet{luo2023steal}) and one backdoor-based (DIAGNOSIS \citep{wang2023diagnosis}). Note that the watermark extraction accuracy and backdoor success rate can be converted into TPR at a certain $\alpha$ through a hypothesis testing process, as described in their original paper \citep{yu2021artificial,wang2023diagnosis}. As a result, we can use TPR in a unified way to compare all methods fairly. More configurations on the baselines and hypothesis testing details can be found in Appendix \ref{app:details}.

\vspace{0.3em}
\noindent\textbf{Implementation Details.} By default, we set the weighting parameters as $\lambda_{1}=1$ and $\lambda_{2}=1$. Following \citep{ruff2018deep}, we set $\nu=0.5$ in our experiments. Following previous practices \citep{li2023black,jia2021entangled}, we set $n=30$ and $m=30$ in Eq. (\ref{eq:kstest}) by default. For all personalization techniques, the training hyperparameters (\eg learning rate, batch size) follow the default setting in their original paper. The generation parameters follow the official default setting provided by HuggingFace. The empirical distributions of $F$ and $G$ are estimated by sending the same prompts for personalized generation to the suspicious model and a benign model (\ie Stable Diffusion v1.5 fine-tuned on the uncoated dataset) and calculating coating scores on the generated images. For each experiment, we repeat the test in Eq. (\ref{eq:kstest}) for 10,000 times, each time with a randomly selected sample set from an image pool of 1,000 generated mimicries, and report the averaged results. More implementation details, including the model structure of {\sc Siren}, are given in Appendix \ref{app:details}.

\begin{figure*}[th]
    \centering
    \begin{subfigure}[b]{0.28\textwidth}
        \centering
        \includegraphics[width=\textwidth]{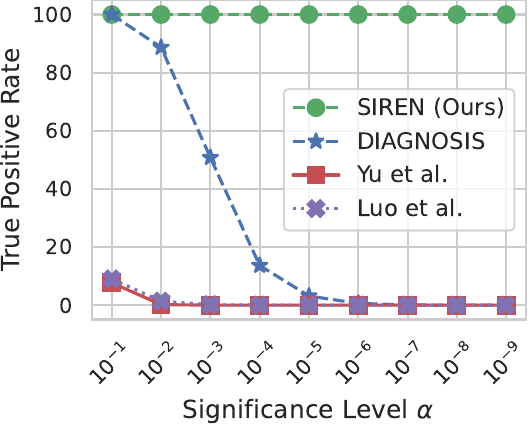}
        \caption{DreamBooth \citep{ruiz2023dreambooth}}
    \end{subfigure}
    \hfill
    \begin{subfigure}[b]{0.28\textwidth}
        \centering
        \includegraphics[width=\textwidth]{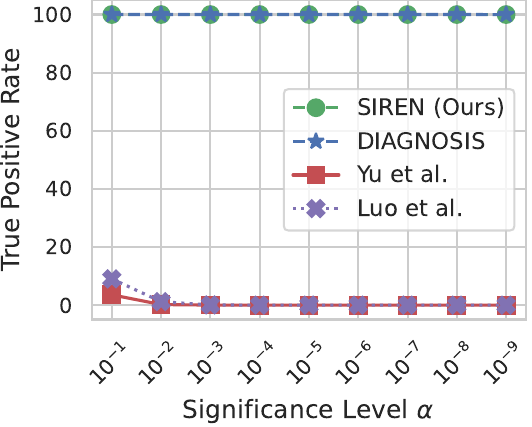}
        \caption{SVDiff \citep{han2023svdiff}}
    \end{subfigure}
    \hfill
    \begin{subfigure}[b]{0.28\textwidth}
        \centering
        \includegraphics[width=\textwidth]{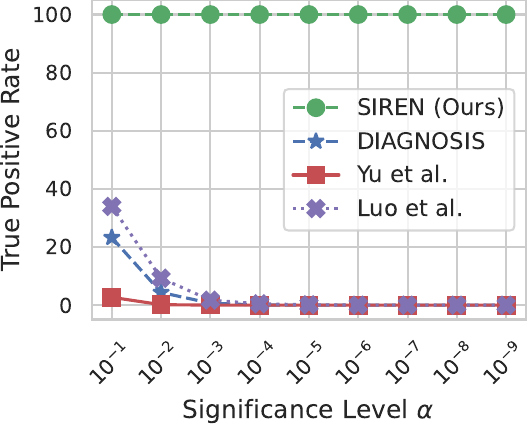}
        \caption{Custom Diffusion \citep{kumari2023multi}}
    \end{subfigure}
    \caption{Effectiveness comparison in the advanced personalization methods. The dataset is Dog \citep{ruiz2023dreambooth}.}
    \label{fig:adv-dog}
    \vspace{-0.5em}
\end{figure*}

\begin{figure*}[th]
    \centering
    \begin{subfigure}[b]{0.28\textwidth}
        \centering
        \includegraphics[width=\textwidth]{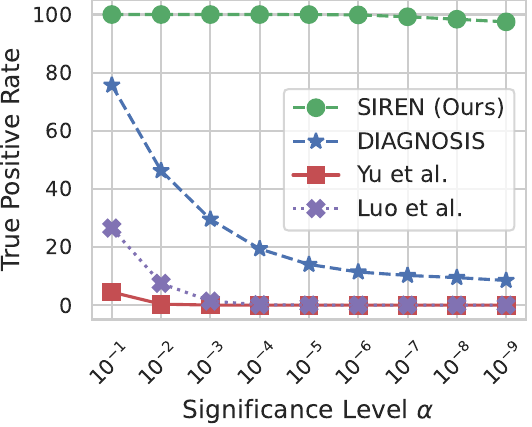}
        \caption{DreamBooth \citep{ruiz2023dreambooth}}
        \label{fig:sub1}
    \end{subfigure}
    \hfill
    \begin{subfigure}[b]{0.28\textwidth}
        \centering
        \includegraphics[width=\textwidth]{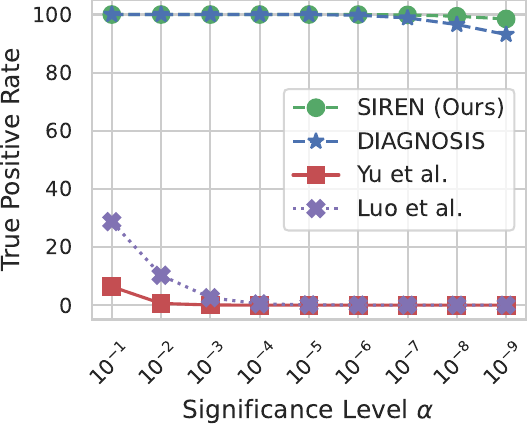}
        \caption{SVDiff \citep{han2023svdiff}}
        \label{fig:sub2}
    \end{subfigure}
    \hfill
    \begin{subfigure}[b]{0.28\textwidth}
        \centering
        \includegraphics[width=\textwidth]{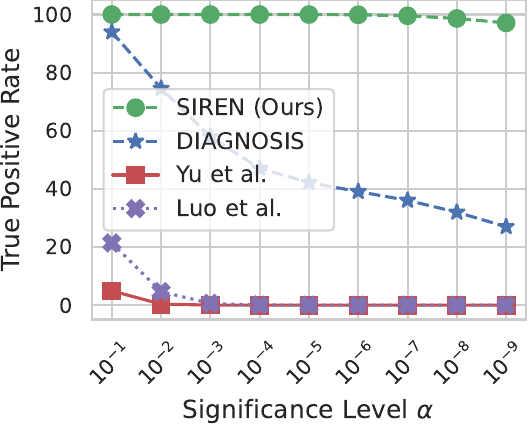}
        \caption{Custom Diffusion \citep{kumari2023multi}}
        \label{fig:sub3}
    \end{subfigure}
    \caption{Effectiveness comparison in the advanced personalization methods. The dataset is WikiArt \citep{saleh2015large}.}
    \label{fig:adv-wikiart}
    \vspace{-1em}
\end{figure*}

\subsection{Effectiveness against Fine-tuning}
We first evaluate the effectiveness of \sys in the standard fine-tuning scenario. We fine-tune the Stable Diffusion v1.5 model using the 4 datasets described in Section \ref{sec:expsetup}. For CelebA-HQ, we use its provided text descriptions as the training prompts. For the other datasets without such text descriptions, we use the BLIP image captioning model \citep{li2023blip} to generate the descriptions as the training prompts. The test-time personalization prompts are set as ``an image of a/an [V]'', where [V] is ``Pokemon'', ``person'', ``artwork'', and ``landscape'' for the corresponding datasets. We train 3 independent models with different random seeds for each experiment and report the averaged results.

Figure \ref{fig:ft-main} shows the evaluation results. Here, the x-axis controls the significance level ($\alpha$) while the effectiveness of each method is presented as the TPR at a certain $\alpha$. Overall, our proposed \sys achieves a TPR of nearly $100\%$ even at $\alpha=10^{-9}$ in all tested datasets. In sharp contrast, the state-of-the-art verification-based defenses either completely fail to effectively detect the unauthorized data usage in personalized diffusion models, or have very fluctuating performances across different datasets. For example, while the backdoor-based method DIAGNOSIS \citep{wang2023diagnosis} performs well on Pokemon and CelebA-HQ, its effectiveness drops quickly on ArtBench and Landscape. On the other hand, the watermark-based methods \citep{yu2021artificial, luo2023steal} completely fail to reach $\text{TPR}>40\%$ in all evaluated cases. The possible reason, as we discussed previously, is that these methods ignore the relevance of injected watermarks/backdoors with the personalization task. For example, DIAGNOSIS's trigger is a warping-based operation that twists the edges of the image subject. These features might be considered by a pre-trained diffusion model as relevant to Pokemon characters or person faces, but may hardly be considered as a feature of artwork or landscape image. This possibly explains why DIAGNOSIS can achieve good results on Pokemon and CelebA-HQ while performing poorly on ArtBench and Landscape.

\vspace{-0.5em}
\subsection{Effectiveness against Advanced Personalization Methods}
\vspace{-0.5em}
We further evaluate the effectiveness of \sys when the model is customized using more advanced personalization methods (\ie DreamBooth \citep{ruiz2023dreambooth}, SVDiff \citep{han2023svdiff}, and Custom Diffusion \citep{kumari2023multi}). We choose two datasets widely used in personalization: Dog \citep{ruiz2023dreambooth} and Wikiart subset \citep{saleh2015large}. For all methods, following \citet{ruiz2023dreambooth}, we set both the user's training prompt and test-time personalization prompt as ``an image of a [V*] [class]'', where [V*] is the personalization pseudo word and [class] is the class noun, automatically acquired as described in \citep{ruiz2023dreambooth}. We note that the test-time personalization prompt (\ie the pseudo word [V*]) is naturally accessible in our threat model: as the attacker deploys the black-box personalized model for profit, he naturally provides [V*] to the users for generating mimicries. Otherwise, the users cannot use this model for personalized generation. Other hyperparameters follow the original settings in their paper. 

The evaluation results for the two datasets are shown in Figure \ref{fig:adv-dog} and Figure \ref{fig:adv-wikiart}. Interestingly, we find that even on the same dataset, the baselines exhibit totally different performances against different personalization methods. For instance, DIAGNOSIS remains effective on SVDiff for both datasets, but it is less effective on DreamBooth and Custom Diffusion. These observations suggest that the baselines are not universal. In contrast, our \sys consistently achieves high effectiveness (TPR) at very low significance levels for all advanced personalization methods and both datasets.

\subsection{Coating Robustness}
In this section, we assess the robustness of \sys under various real-world scenarios.

\vspace{0.3em}
\noindent\textbf{The protector/infringer uses different models/prompts.} Recall that the $\mathcal{L}_{\text{learn}}$ term in Eq. (\ref{eq:learn}) is calculated using the surrogate model and surrogate prompt. We investigate whether \sys remains effective when there exists different degrees of divergence between the infringer's actual model and surrogate model. Specifically, we select 4 state-of-the-art opensource text-to-image diffusion models for transferability evaluation: Stable Diffusion v2.1 \citep{stabilityai_stable_diffusion_2_1}, Kandinsky 2.2 \citep{razzhigaev2023kandinsky}, Latent Consistency Models \citep{luo2023latent}, and VQ Diffusion \citep{gu2022vector}. Stable Diffusion v2.1 has the same architecture with the surrogate model while trained with different datasets and settings, and other models are totally different from the surrogate model in terms of architecture, training set, and hyperparameters. For training prompts, we use the prompt generated by three different state-of-the-art image captioning models: BLIP \citep{li2023blip}, LLaVA \citep{liu2023llava}, and PaLI \citep{chen2023pali}. 

As shown in Table \ref{tab:transferability}, \sys exhibits very high transferability across all evaluated models and training prompts, achieving a TPR of $100\%$. This is not surprising -- previous works have shown that the ``semantic perturbations'' learned from Stable Diffusion models have high transferability \citep{liu2024metacloak}. Moreover, \sys also has good transferability across training prompts generated by different captioning models.

\begin{table}[t]
    \centering
    \caption{Transferability of \sys across different diffusion models and training prompts. The reported metric is the TPR at $\alpha=10^{-9}$.}
    \label{tab:transferability}
    \footnotesize
    \vspace*{-0.2em}
    \setlength{\tabcolsep}{6pt}
    \resizebox{0.97\linewidth}{!}{
        \begin{tabular}{ccccc}
            \toprule
            \multirow{3}{*}{\bf{Dataset}} & \multirow{3}{*}{\bf{Model}} & \multicolumn{3}{c}{\bf{Training Prompt Generator}} \\
            \cmidrule(lr){3-5}
            & & \bf{BLIP} & \bf{LLaVA} & \bf{PaLI} \\
            \midrule
            \multirow{4}{*}{Pokemon} & Stable Diffusion v2.1 \citep{stabilityai_stable_diffusion_2_1} & $100\%$ & $100\%$ & $100\%$ \\
            & Kandinsky 2.2 \citep{razzhigaev2023kandinsky} & $100\%$ & $100\%$ & $100\%$ \\
            & Latent Consistency Models \citep{luo2023latent}& $100\%$ & $100\%$ & $100\%$ \\
            & VQ Diffusion \citep{gu2022vector}& $100\%$ & $100\%$ & $100\%$ \\
            \midrule
            \multirow{4}{*}{CelebA-HQ} & Stable Diffusion v2.1 \citep{stabilityai_stable_diffusion_2_1} & $100\%$ & $100\%$ & $100\%$ \\
            & Kandinsky 2.2 \citep{razzhigaev2023kandinsky} & $100\%$ & $100\%$ & $100\%$ \\
            & Latent Consistency Models \citep{luo2023latent}& $100\%$ & $100\%$ & $100\%$ \\
            & VQ Diffusion \citep{gu2022vector}& $100\%$ & $100\%$ & $100\%$ \\
            \bottomrule
        \end{tabular}
    }
    \vspace{-2em}
\end{table}

\vspace{0.3em}
\noindent\textbf{The training set consists of both coated and clean images.} Next, we consider another practical scenario where the training set collected by the infringer includes both coated and clean images. This is realistic because an infringer may collect the dataset from multiple sources, while the user's images may only be part of it. Note that as the ratio of coated images over the training set decreases, the final mimicries would be much less similar to the user's images \citep{wang2023diagnosis,shan2023glaze}.

\begin{figure}[t]
    \centering
    \begin{subfigure}[b]{0.23\textwidth}
        \centering
        \includegraphics[width=\textwidth]{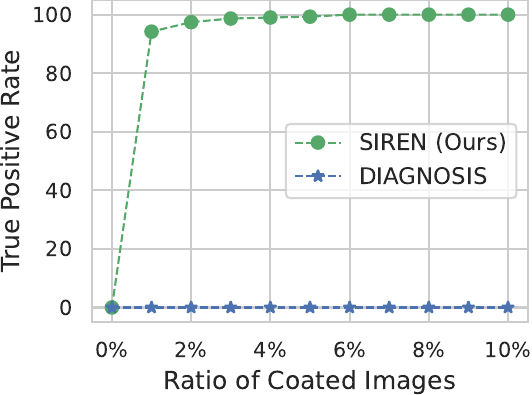}
        \caption{Pokemon  ($\alpha=10^{-4}$)}
    \end{subfigure}
    \hfill
    \begin{subfigure}[b]{0.23\textwidth}
        \centering
        \includegraphics[width=\textwidth]{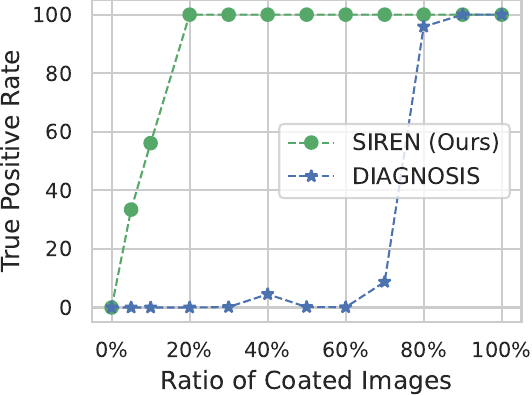}
        \caption{Pokemon  ($\alpha=10^{-9}$)}
    \end{subfigure}
    \begin{subfigure}[b]{0.23\textwidth}
        \centering
        \includegraphics[width=\textwidth]{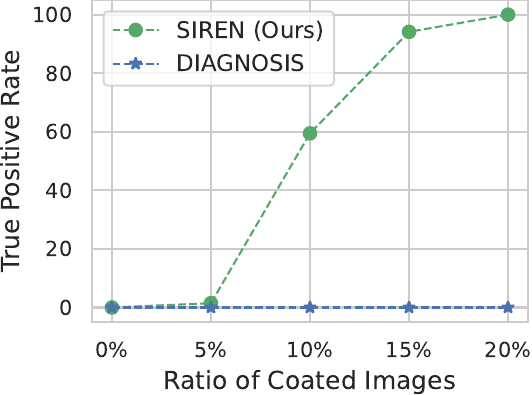}
        \caption{CelebA-HQ  ($\alpha=10^{-4}$)}
    \end{subfigure}
    \hfill
    \begin{subfigure}[b]{0.23\textwidth}
        \centering
        \includegraphics[width=\textwidth]{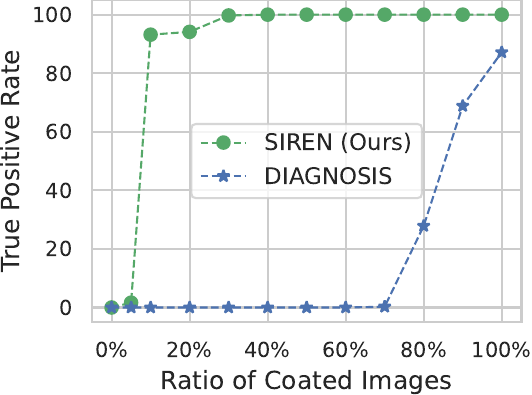}
        \caption{CelebA-HQ ($\alpha=10^{-9}$)}
    \end{subfigure}
    \caption{Protection robustness when the training dataset consists of both coated images and uncoated images.}
    \label{fig:coating-ratio}
\end{figure}

We evaluate the robustness of \sys and compare it to DIAGNOSIS in this setting. As can be seen from Figure \ref{fig:coating-ratio}, \sys is still highly effective and significantly outperforms DIAGNOSIS: on both datasets, \sys almost achieves a TPR of $100\%$ at $\alpha=10^{-9}$ when the ratio of coated images exceeds $20\%$. On the Pokemon dataset \sys even reaches a surprisingly high TPR of $94.2\%$ at $\alpha=10^{-4}$ when the coated dataset only consists $1\%$ of the entire training set.

One may note that \sys is more effective on Pokemon than on CelebA-HQ. One possible reason is that the model's training dataset (\ie LAION-5B) already includes the entire CelebA-HQ dataset, while Pokemon images are not included. Consequently, since the base model has already seen CelebA-HQ during pre-training, it tends to learn less new knowledge when trained on it again. As a result, the coating generated by \sys is less effective, especially at low coating ratios. This phenomenon is also observed for DIAGNOSIS and adversarial-based protections \citep{shan2023glaze}. However, we believe this is not a significant issue: the primary target of both personalization learning and our \sys are those images that have not been seen by diffusion models during pre-training (\ie the new concepts). In this scenario, our method is still highly effective at very low coating ratios.

\vspace{0.3em}
\noindent\textbf{Other Experiments.} We conduct some additional experiments, which explore the effectiveness \sys when (1) the mimicries undergo further transformations, (2) data size is small, (3) the model is further modified, and (4) the infringer uses different generation prompts and hyperparameters. Overall, \sys is highly effective in these scenarios. More results and analyses can be found in Appendix \ref{app:exp}.

\subsection{Impact on Image Quality}
In this section, we investigate the impact of \sys on image quality. Specifically, we assess ({1}) whether patching the coating degrades the visual quality of the training set images; and ({2}) whether the quality of  mimicries generated by models personalized with coated images degrades. We first evaluate the performance quantitatively using automatic metrics, then we include a human evaluation to fully understand the impact of \sys on human-perceived quality.  

\begin{figure}[t]
    \centering
    \includegraphics[width=\linewidth]{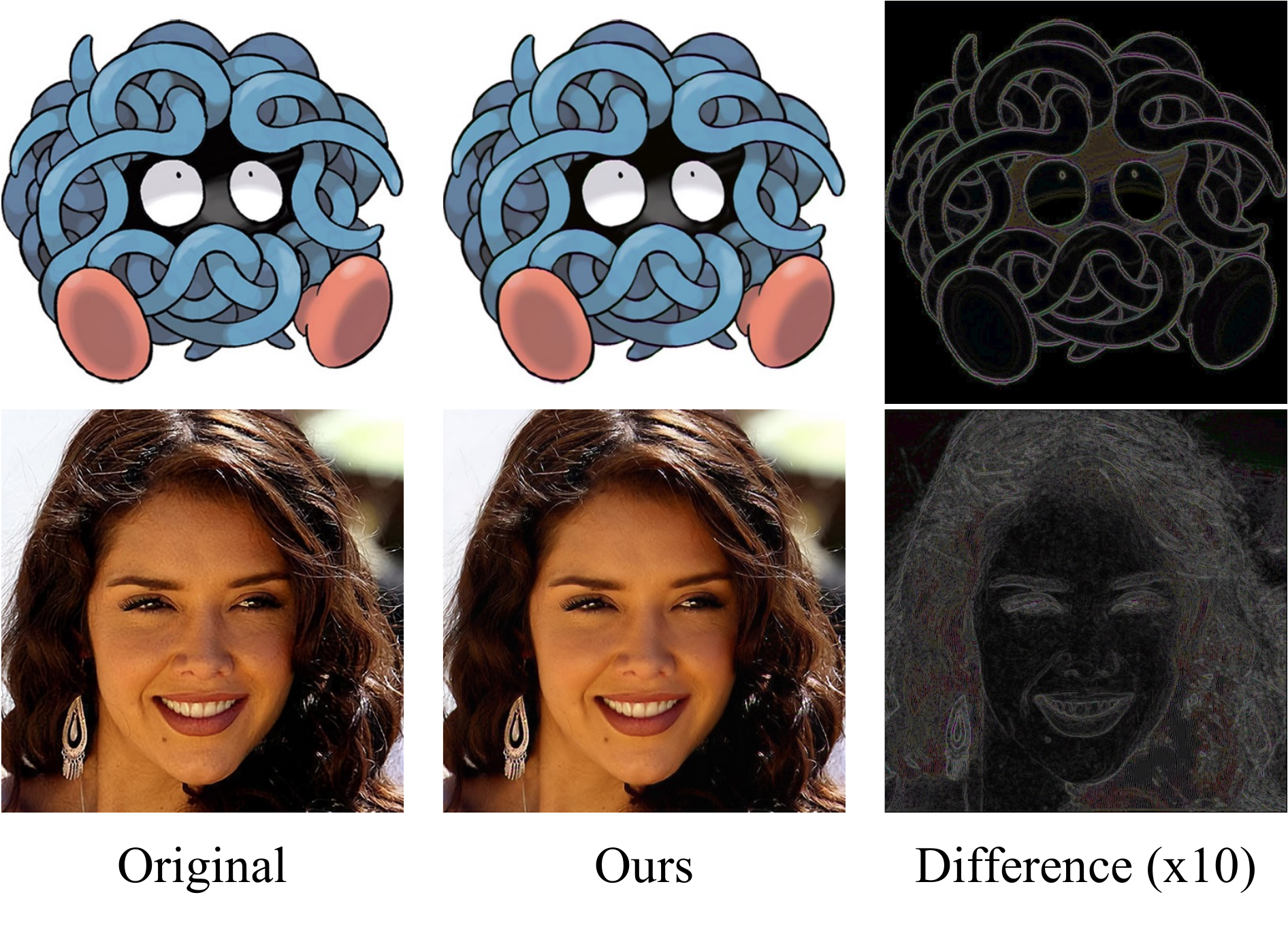}
    \vspace{-1.5em}
    \caption{\small The original images and their coated version.}
    \vspace{-0.5em}
    \label{fig:coated_images}
\end{figure}

\vspace{0.3em}
\noindent\textbf{Qualitative and Quantitative Evaluations.} As can be seen from the quantitative results in Table \ref{tab:datasetquality}, the coating has overall a high PSNR, SSIM, and low LPIPS values on the datasets evaluated. We also provide some qualitative results in Figure \ref{fig:coated_images}, which show that the perturbations generated by \sys are generally imperceptible to human observers.
\begin{table}[t]
    \centering
    \caption{Impact on visual quality of the coated dataset.}
    \label{tab:datasetquality}
    \footnotesize
    \setlength{\tabcolsep}{6pt}
    \resizebox{0.8\linewidth}{!}{
        \begin{tabular}{cccccc}
            \toprule
            \bf{Dataset} & \bf{PSNR $\uparrow$} & \bf{SSIM $\uparrow$} & \bf{LPIPS $\downarrow$}  \\
            \midrule
            {Pokemon}
            &  {40.51} & {0.993} & {0.0038} \\
            \midrule
            {CelebA-HQ}
            & 38.17 & 0.951 & 0.0453 \\
            \midrule
            {Dog}
            & 40.52 & 0.972 & 0.0268 \\
            \bottomrule
        \end{tabular}
    }
    \vspace{-1em}
\end{table}

\begin{table}[t]
    \centering
    \caption{\small Impact on generation quality. Original means fine-tune the model using the original uncoated dataset, while \sys indicates to fine-tuning the model using the dataset coated by {\sc Siren}.} 
    \label{tab:generation_quality}
    \footnotesize
    \vspace*{-0.2em}
    \setlength{\tabcolsep}{6pt}
    \resizebox{0.96\linewidth}{!}{
        \begin{tabular}{ccccc}
            \toprule
            \bf{Dataset} & \bf{Setting} & \bf{FID $\downarrow$} & \bf{CLIP Score $\uparrow$} & \bf{DINO Score $\uparrow$} \\
            \midrule
            \multirow{2}{*}{Pokemon} & Original & 104.57 & 0.816 & 0.701 \\
            & \sys (Ours) & 103.26 & 0.828 & 0.709\\
            \midrule
            \multirow{2}{*}{CelebA-HQ} & Original & 63.57 & 0.574 & 0.605\\
            & \sys (Ours) & 59.07  & 0.576 & 0.612 \\
            \midrule
            \multirow{2}{*}{Dog} & Original & 58.00 & 0.910 & 0.835 \\
            & \sys (Ours) & 59.36 & 0.908 & 0.829 \\
            \bottomrule
        \end{tabular}
    }
    \vspace{-1em}
\end{table}

\begin{figure}[t]
    \centering
    \includegraphics[width=\linewidth]{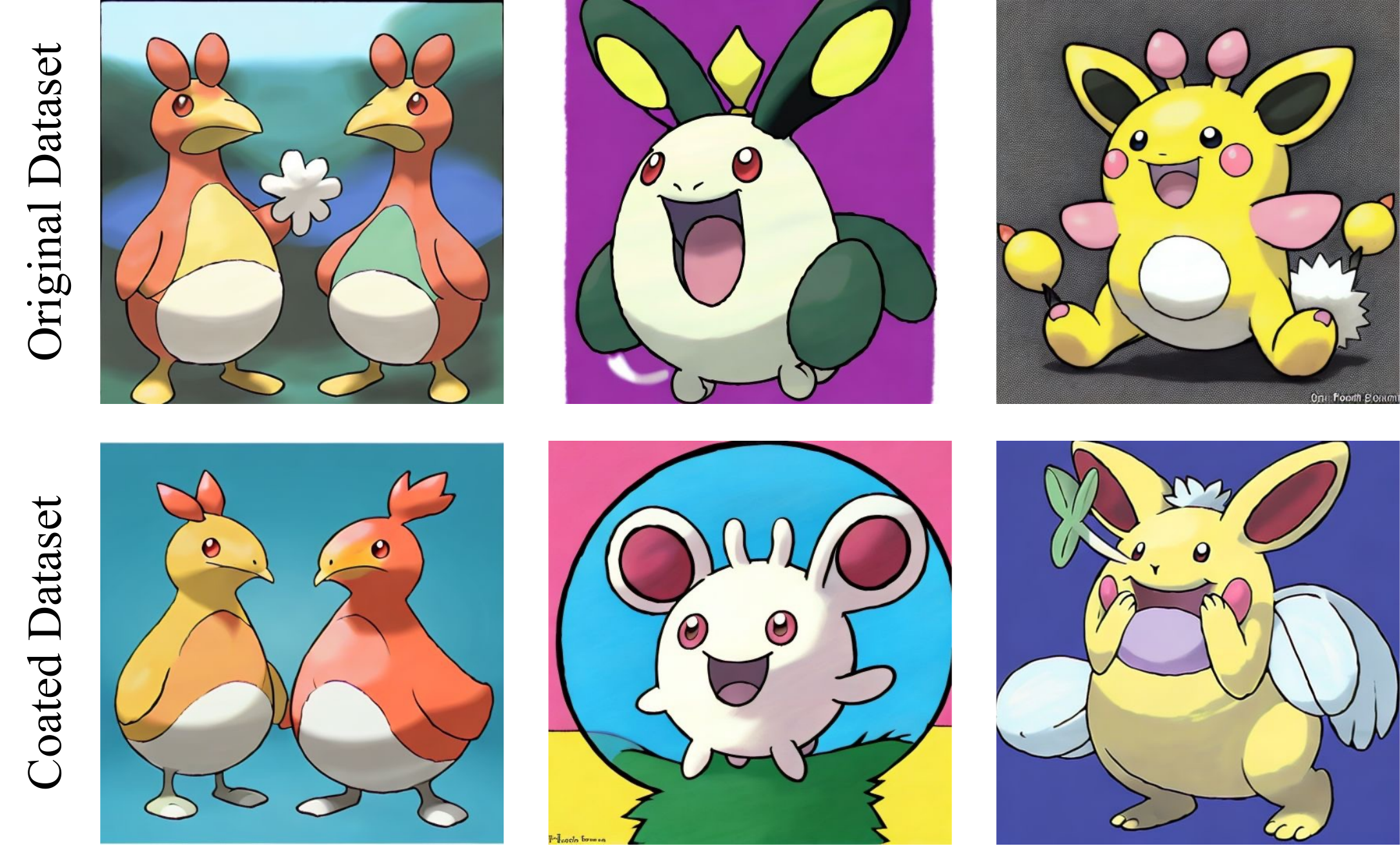}
    \caption{\small Generated images with original dataset and the dataset coated by {\sc Siren}.}
    \label{fig:generated_images}
\end{figure}

We then evaluate {\sc Siren}'s impact on the generation quality of the personalized model. The results in Table \ref{tab:generation_quality} show that the impact of \sys on generation quality is small, as evidenced by a small difference of all metrics. Some generation examples are provided in Figure \ref{fig:generated_images} (Appendix).

\begin{figure}[t]
    \centering
    \begin{subfigure}{0.23\textwidth}
        \centering
        \includegraphics[width=\linewidth]{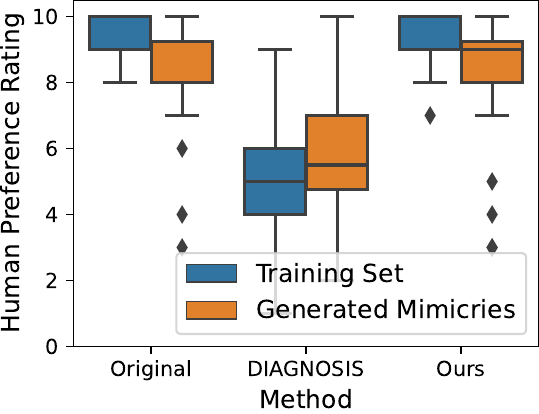}
        \caption{Pokemon}
    \end{subfigure}
    \hfill
    \begin{subfigure}{0.23\textwidth}
        \centering
        \includegraphics[width=\linewidth]{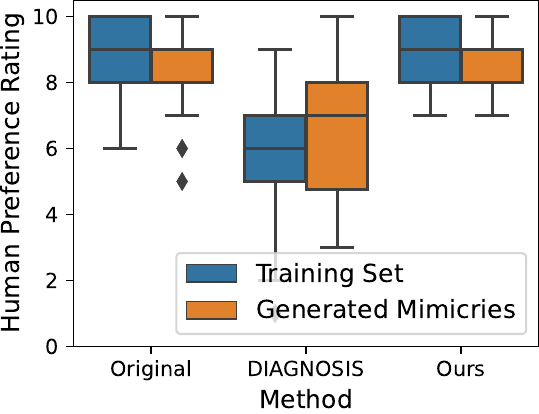}
        \caption{CelebA-HQ}
    \end{subfigure}
    \caption{\small Human preference study results. \sys only causes a very small impact on both training dataset and generated mimicries, while DIAGNOSIS produces much more visible artifacts.}
    \label{fig:human} 
    \vspace{-0.5em}
\end{figure}

\vspace{0.3em}
\noindent\textbf{Human Preference Study.} Finally, we assess the impact of \sys on image quality through a human preference study on Pokemon and CelebA-HQ. We compare our method with DIAGNOSIS as it is the most effective baseline. For each dataset, we randomly choose 6 training images and protect them by DIAGNOSIS and {\sc Siren}, respectively. Then, we randomly select 6 images generated by personalized models trained on unprotected, DIAGNOSIS-protected, and {\sc Siren}-protected datasets, respectively. We then prepare a survey with 24 questions, each displaying three images in random order (original, DIAGNOSIS, and {\sc Siren}). Participants are asked to rate each image based on quality and naturalness (see more details and a sample question in Appendix \ref{app:details}). The rating, which we refer to as human preference rating (HPR), ranges from 1$\sim$10, where 7$\sim$10 indicates very good quality and high naturalness, 4$\sim$6 indicates some low-quality details and visible, unnatural artifacts, and 1$\sim$3 indicates very low quality and very unnatural appearance. For generated images, we additionally ask the participants to consider the similarity to the training dataset. The study is performed with 32 volunteer university students and faculties aged between 20-33, with $32\times24\times3=2208$ answers in total. The whole procedure has been reviewed and approved by our school’s IRB, whose process is similar to the exempt review in the US, since the study is considered as ``minimal risk'' by IRB staff. The results are summarized in Figure \ref{fig:human}. To summarize, \sys only causes a small impact on both training dataset and generated mimicries, and significantly outperforms DIAGNOSIS in all evaluated settings.

\subsection{Real-world Case Studies}

We evaluate the effectiveness of \sys on two real-world personalization-as-a-service platforms, \ie Replicate \citep{replicate2024} and Scenario \citep{scenario2024}. These platforms provide online personalized model training and  sharing services. The user only needs to upload the reference dataset, and the platform will automatically train the personalized model and return the API for mimicry generation. In this scenario, the base model, detailed personalization algorithm and configurations, training and generation prompts, as well as image pre-processing and post-processing methods are all controlled by the service provider and unknown to {\sc Siren}, making it more challenging compared to local training.

We feed Pokemon and CelebA-HQ coated by \sys to both services and ask them to train a personalized model for each dataset. As shown in Table \ref{tab:realworld}, \sys is highly effective, reaching a $\text{TPR}=100\%$ at $\alpha=10^{-9}$ in all evaluated cases. Note that the FID is slightly higher, possibly because these platforms use fewer iteration steps than local training.

\begin{table}[ht]
    \centering
    \caption{\small Performance of {\sc Siren} in real-world personalization-as-a-service services. $\alpha$ is set to $10^{-9}$ in this experiment.}
    \label{tab:realworld}
    \footnotesize
    \vspace*{-0.2em}
    \setlength{\tabcolsep}{6pt}
    \resizebox{0.7\linewidth}{!}{
        \begin{tabular}{cccc}
            \toprule
            \bf{Dataset} & {\bf{Service}} &  \bf{TPR $\uparrow$} & \bf{FID $\downarrow$}\\
            \midrule
            \multirow{2}{*}{{Pokemon}} & Replicate & $100\%$ & $164.42$  \\
            & Scenario & $100\%$ & $179.77$  \\
            \midrule
            \multirow{2}{*}{{CelebA-HQ}} & Replicate & $100\%$ & $124.35$  \\
            & Scenario & $100\%$ & $133.27$  \\
            \bottomrule
        \end{tabular}
    }
    \vspace{-1.5em}
\end{table}

\subsection{\sys against Potential Countermeasures}
We consider several potential countermeasures the infringer might take and verify whether they can reduce the effectiveness of {\sc Siren}. Based on the infringer's goal, an attack is considered successful if it can evade detection (\eg degrading the TPR to very low) while ensuring the generation quality of the model is not severely harmed. 

\vspace{0.3em}
\noindent\textbf{Outlier detection}.  This technique detects abnormal data points that are far from the main distribution. The attacker may use it to identify coated images. To verify whether outlier detection can robustly detect {\sc Siren}'s coating, we use the state-of-the-art outlier detection model  \citep{reiss2021panda}. We split Pokemon into a training set and test set with the ratio of 8:2. Then, we train the model on the uncoated Pokemon training set and use it to detect whether the coated/clean version of the Pokemon validation set are outliers. The results (AUC=$51\%$, Recall=$52\%$) indicate that outlier detection is not successful in effectively identifying {\sc Siren}'s coating.

\begin{figure}[t]
    \centering
    \includegraphics[width=\linewidth]{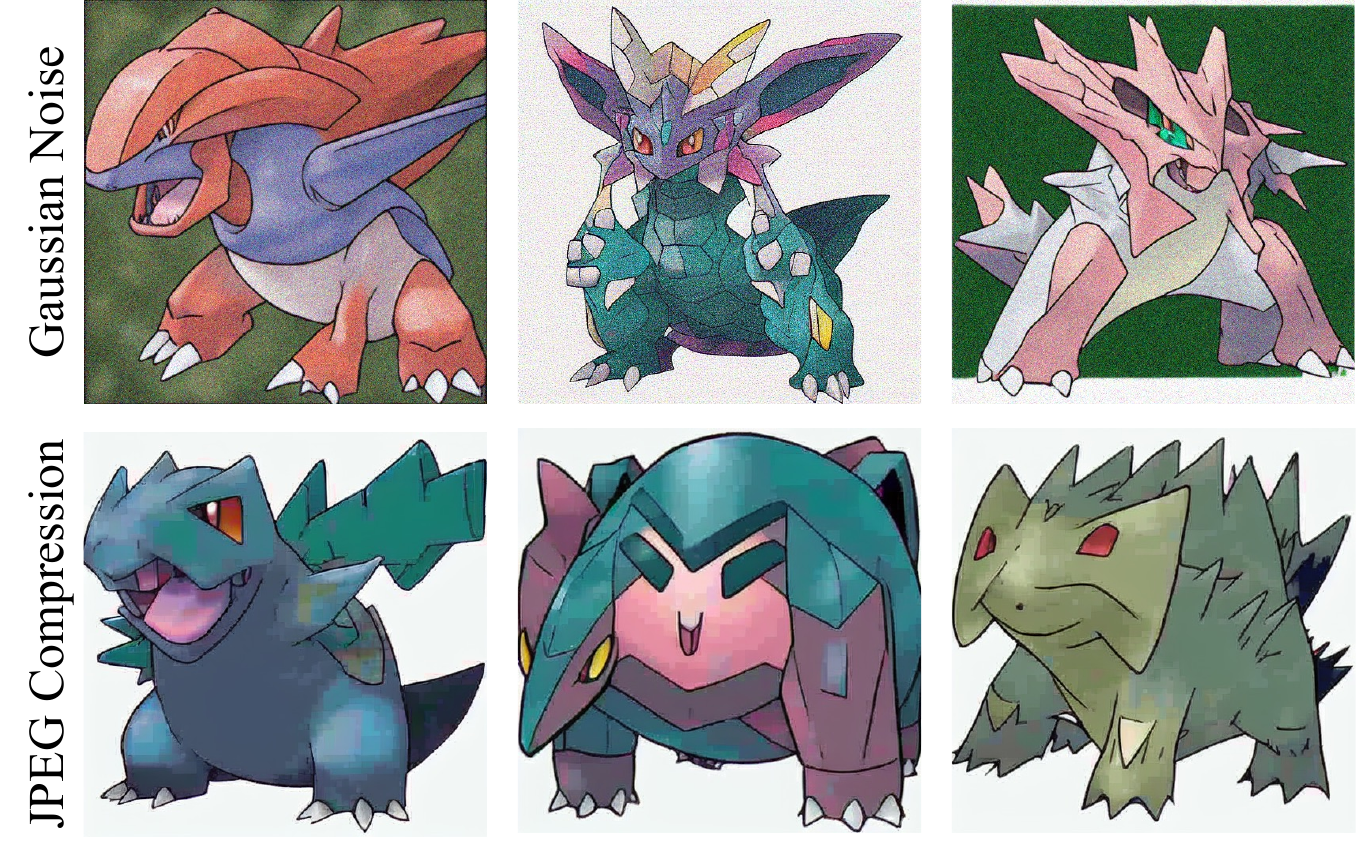}
    \caption{\small Images generated by the personalized model when the dataset coated by \sys undergoes training-time augmentations. When the training set is added with Gaussian noise ($\sigma=0.1$) or is JPEG compressed (factor=$40$), the generated images already contain obvious artifacts (zoom in for a better view). However,  {\sc Siren} can still achieve high performance at this level.}
    \label{fig:adaptive_train}
    \vspace{-1em}
\end{figure}

\vspace{0.3em}
\noindent\textbf{Training-time augmentation}. This approach is widely used to eliminate small image perturbations \citep{shan2023glaze}. We try 2 types of augmentations on the training images: adding Gaussian noise ($\sigma=0.1$) and JPEG compression (factor=$40$). Notably, these augmentations have already harmed the generation quality of the model: models trained on compressed/noisy images also learn to replicate similar artifacts in their mimicries, as evidenced in Figure \ref{fig:adaptive_train} (Appendix). Our human evaluation averaged over 10 generations also indicates that human observers can easily see obvious artifacts on the mimicries (HPR=3.5 on Gaussian noise and 3.7 on JPEG compression). However, \sys is still effective in this scenario: it achieves a TPR of $98.7\%$ and $100\%$ when setting $\alpha$ to $10^{-9}$ on Gaussian noise and JPEG compression, respectively. Overall, the attacker cannot easily bypass \sys using straightforward training-time augmentations without harming the quality of the mimicries.

\begin{figure}[h]
    \centering
    \vspace{-0.5em}
    \includegraphics[width=0.75\linewidth]{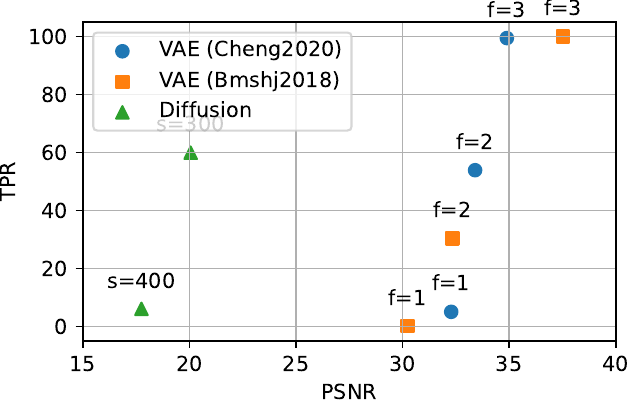}
    \caption{\small Results of purification attacks. $\alpha$ is set to $10^{-9}$. f (factor) and  s (steps) indicate purification strengths for VAE and diffusion models, respectively. Lower f and higher s bring more successful attack performance yet worse image quality. The PSNR is calculated on the original image and its purified version.}
    \vspace{-0.5em}
    \label{fig:purification-psnr}
\end{figure}

\vspace{0.3em}
\noindent\textbf{Post-generation purification}. This technique alters the generated mimicries via regeneration-based perturbation purification. The general idea is to first ``destroy'' the original, perturbed image, and then reconstruct a ``clean'' version of it using a generative model, such as VAE \citep{cheng2020learned} or diffusion models \citep{xu2023invisible}. It has strong (or even certified) performance in removing adversarial perturbations or image watermarks. We conduct experiments following the protocol in \citep{xu2023invisible} with different generative models and different levels of attack strengths, and the results are shown in Figure \ref{fig:purification-psnr}. Overall, the attacker can successfully evade our detection when the distortion is sufficiently large: he can replace the generated image with a completely different clean image to evade detection anyway. Empirically, we observe this solution brings unnatural artifacts or blurry areas in the mimicries, significantly degrading the quality of the mimicries. For example, Bmshj2020 (factor=2) can effectively reduce {\sc Siren}'s TPR to 91.74\% ($\alpha=10^{-6}$) and 30\% ($\alpha=10^{-9}$). However, it only gains an averaged HPR of $2.2$, and over $90\%$ of the participants gave ratings of less than $3$ (\ie very low quality). The more successful diffusion attack (step=400) only gains an averaged human rating of $1.1$, and over $90\%$ of human testers rate it as $1$ (lowest score). This is possibly because the learnability loss makes \sys absorbed as an inherent semantic feature of the target class, making it hard to remove unless the semantics (or quality) of the mimicries are destroyed. Some example images that can successfully evade our detection are given in Figure \ref{fig:purification-results} (Appendix). Another watermark removal attack \citep{JiangZG23} is discussed in Appendix \ref{app:discussion}.

\begin{table}[t]
    \centering
    \caption{\small Performance of {\sc Siren} under ABL. $\alpha$ is set to $10^{-4}$.}
    \label{tab:abl}
    \footnotesize
    \vspace*{-0.2em}
    \setlength{\tabcolsep}{6pt}
    \resizebox{0.6\linewidth}{!}{
        \begin{tabular}{cccc}
            \toprule
            \multirow{2}{*}{\bf{Setting}} & \multicolumn{3}{c}{\bf{Coating Rate}} \\
             & {\bf{5\%}} &  \bf{10\%} & \bf{15\%}\\
            \midrule
            Standard Training & 100\% & 100\% & 100\% \\
            w/ ABL & 99.83\% & 100\% & 100\%\\
            \bottomrule
        \end{tabular}
    }
    \vspace{-2em}
\end{table}

\vspace{0.3em}
\noindent\textbf{Loss-based filtering and unlearning.} We design an adaptive attack according to the knowledge of {\sc Siren}, which is based on the intuition that the coating optimized by $\mathcal{L}_{\text{learn}}$ might be more ``attractive'' than other features, similar to semantic backdoor triggers \citep{zeng2023narcissus}, so it might be learned faster than other features. To this end, we leverage the idea of ABL \citep{li2021anti} to implement a loss-based filtering and unlearning attack. ABL is a training-time backdoor mitigation method that leverages similar observations on neural backdoor triggers (\ie the backdoor task is usually learned faster than the normal one). Building upon this fact, ABL first filters suspicious samples according to the loss, and then uses gradient ascent to unlearn the suspicious features (\ie the trigger). We extend ABL to the diffusion model training setting (more details are in Appendix \ref{app:details}) and test whether it can successfully evade {\sc Siren}. Specifically, we coat the Pokemon dataset with different coating rates and use ABL to detect and unlearn the coating. The filter rate of ABL is set to 5\%. As shown in Table \ref{tab:abl}, ABL has only limited effect in bypassing {\sc Siren}. We also check the filter results of ABL and find only 4 out of 83 coated images are filtered by it when the coating rate is 10\%, while the other 38 filtered images are all clean. This suggests {\sc Siren}'s coating would be considered similar to the other features with a similar loss scale, thus making this strategy less effective.

\begin{table}[h]
\vspace{-0.5em}
    \centering
    \footnotesize
    \setlength{\tabcolsep}{4pt}
    \caption{{Results when attacker learns to uncoat with auxiliary datasets.} PSNR is calculated between the original mimicries and their purified version. $\alpha$ is set to $10^{-9}$.}
        \centering
         \scalebox{1}{
         \begin{tabular}{c c c}
         \toprule
               \textbf{Auxiliary Dataset}  & \textbf{PSNR} $\uparrow$ & \textbf{TPR} $\uparrow$ \\ 
        \midrule
               Anime-Chibi  & 20.20 & 100 \\
               Pokemon*  & 24.87 & 100 \\
        \bottomrule
        \end{tabular}
        }
    \label{tab:auxiliary-dataset}
    \begin{flushleft}
    \footnotesize{* We split the Pokemon training set into two non-overlapping subsets (in a ratio of 1:1). We assume the user owns the first half and the infringer uses the second half to learn the mapping and conduct the attack.}
    \end{flushleft}
    \vspace{-1.5em}
\end{table}

\vspace{0.3em}
\noindent\textbf{Learning to uncoat with auxiliary datasets.} Finally, we consider a scenario where the infringer has a clean auxiliary dataset $\{x_\mathcal{A}\}$ with a similar (or even same) distribution to the user's data $\{x\}$. In this scenario, the infringer can ask the platform to train a coating generator $\mathcal{G}_\mathcal{A}$ and coat his/her images. With these coated images and the original ones, the infringer can learn a mapping $\mathcal{M}:x_\mathcal{A}+\mathcal{G}_\mathcal{A}(x_\mathcal{A}) \rightarrow x_\mathcal{A}$ that ``uncoats'' a given image (\ie inputs a coated image and outputs a clean one). Then, the infringer can leverage $\mathcal{M}$ to conduct a transfer attack on the generated mimicries that contain the user's coatings. We test \sys using the Pokemon dataset under two auxiliary dataset settings: Anime-Chibi \citep{phimsiri2020animechibi} dataset which has a similar distribution to Pokemon, and a subset of Pokemon which has exactly the same distribution. We train a UNet \citep{unet}, an encoder-decoder model for 500 epochs to learn $\mathcal{M}$. The results shown in Table \ref{tab:auxiliary-dataset} prove {\sc Siren}'s resistance to this attack. We speculate that it is because coatings are sample-specific and highly dependent on the training set. Thus, the trained mapping has low transferability on unseen coated images. 

\subsection{Ablation Study}
\label{sec:abl}

In this section, we conduct ablation studies to verify the effectiveness of our each component. We also conduct a hyperparameter analysis in Figure \ref{fig:ablation} (Appendix \ref{app:exp}). 

\vspace{0.3em}
\noindent\textbf{Learnability Loss and Perceptual Constraint.} Learnability loss is the key component for \sys to boost performance, while perceptual constraint helps improve visual performance. As can be seen from Table \ref{tab:ablation-component}, $\mathcal{L}_\text{learn}$ can boost the verification performance, indicated by a large improvement of TPR. However, it also slightly degrades the image quality. $\mathcal{L}_\text{percept}$ serves as a good compensation for image quality, as indicated by both higher PSNR and FID.

\begin{table}[!t]
    \centering
    \footnotesize
    \setlength{\tabcolsep}{4pt}
    \caption{{Ablation study on learnability loss and perceptual constraint.} $\ell_{\infty}$ is a baseline where the coating is directly generated under $\ell_{\infty}$ constraint, while $\mathcal{G}$ indicates using our generator training with an MSE loss on images. The dataset is Pokemon and the model is Stable Diffusion v1.5. $\alpha$ is set to $10^{-9}$ when evaluating TPR. }
        \centering
         \scalebox{1}{
         \begin{tabular}{l | c c c}
         \toprule
               \textbf{Configuration}  & \textbf{PSNR} $\uparrow$ & \textbf{FID} $\downarrow$ & \textbf{TPR} $\uparrow$ \\ 
        \hline
               $\ell_{\infty}$ & 35.04 & 128.67 & 0.40 \\
               \hline
               $\mathcal{G}$ & 39.64 & 105.96 & 0.83  \\
               $\mathcal{G} + \mathcal{L}_{\text{learn}}$ & 39.07 & 107.84 & 100 \\
               $\mathcal{G} + \mathcal{L}_{\text{learn}} + \mathcal{L}_{\text{precept}}$ & {40.51} & {103.98} & {\color{white}0}{100}{\color{white}0} \\
        \bottomrule
        \end{tabular}
        }
    \label{tab:ablation-component}
\end{table}

\begin{table}[!t]
    \centering
    \footnotesize
    \setlength{\tabcolsep}{4pt}
    \caption{{Ablation study on binary and hypersphere classification.} The FPR ($\alpha = 10^{-9}$) is evaluated with two different significance level thresholds ($\alpha = 10^{-9}$ and $\alpha = 10^{-14}$).}
        \centering
         \scalebox{0.95}{
         \begin{tabular}{l | c c c}
         \toprule
               \textbf{Configuration}  & \textbf{TPR} $\uparrow$ & \textbf{FPR} ($\alpha=10^{-9}$) $\downarrow$ & \textbf{FPR} ($\alpha=10^{-14}$) $\downarrow$ \\ 
        \midrule
               Binary & 100 & 100 & 96.05 \\
               Hypersphere & 100 & 0 & 0 \\
        \bottomrule
        \end{tabular}
        }
    \label{tab:classification-results}
    \vspace{-1em}
\end{table}

\vspace{0.3em}
\noindent\textbf{Hypersphere Classification.} As discussed in Section \ref{sec:hypersphere-classification}, a common and intuitive practice to detect the coating is to jointly train a coating detector with standard cross-entropy, and this approach may be biased by the incomplete distribution of negative data. To verify this, we conduct experiments on a unseen benign image dataset, \ie COCO validation set. As shown in Table \ref{tab:classification-results}, using binary classification, the extractor has a high false accusation rate on benign data, as indicated by a high FPR even in very low $\alpha$ regime. In contrast, our hypersphere classification focus on positive data and will be less biased, with a significantly lower FPR.

\begin{figure}[t]
    \centering
    \begin{subfigure}{0.23\textwidth}
        \centering
        \includegraphics[width=\linewidth]{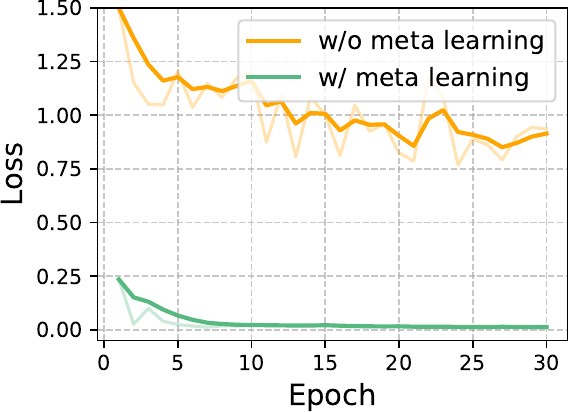}
        \caption{Pokemon}
    \end{subfigure}
    \hfill
    \begin{subfigure}{0.23\textwidth}
        \centering
        \includegraphics[width=\linewidth]{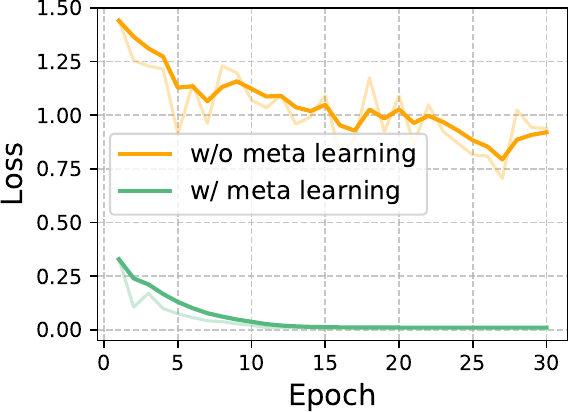}
        \caption{CelebA-HQ}
    \end{subfigure}    
    \vspace{-0.3em}
    \caption{\small Effectiveness of meta learning.}
    \label{fig:meta}
    \vspace{-1.5em}
\end{figure}

\vspace{0.3em}
\noindent\textbf{Meta-learning.} As shown in Figure \ref{fig:meta}, on both datasets, meta-learning provides good initial weights (starting points) compared with random initialization, which helps \sys to converge faster and smoother in future training. 

\section{Conclusion}
This paper introduces {\sc Siren}, a novel methodology for reliable data usage verification in black-box personalized text-to-image diffusion models. \sys enhances the learnability of the coatings by optimizing it to be a feature relevant to personalized learning. We further propose several techniques to improve the stealthiness, effectiveness, and efficiency of {\sc Siren}. We evaluate \sys through extensive experiments and real-world scenarios. We also demonstrate its robustness against different potential countermeasures. 

\vspace{0.3em}
\noindent\textbf{Limitations.} Our \sys still has the following limitations, which we aim to address in future work. First, its detection result can only indicate that the suspicious model is possibly trained on the protected dataset, but cannot imply the IP of this model/dataset totally belongs to the accuser. In fact, the very concept of IP infringement becomes difficult to define strictly from the legal perspective \citep{zirpoli2023generative}, due to the involvement of multiple parties throughout the process and the increasingly blurred boundaries of authorship in the AIGC era. As such, we hope \sys to serve as a valuable reference, rather than definite conclusions. Second, our method cannot detect data misuse if the suspicious model does not accept public users' queries for generating mimicries. However, this also limits the spread of the model. It also cannot detect the unauthorized usage of the datasets whose uncoated versions are previously published online, since infringers can simply use the uncoated dataset to personalize the model.  In such circumstances, the data user may need to cooperate with the model trainer or dataset provider to prevent unauthorized data usage. Finally, while we designed and evaluated several countermeasures, security is an evolving game, and future stronger attacks that can bypass \sys may arise. Designing stronger adaptive attacks and defending \sys against them would be very interesting and meaningful for future work.

\section*{Acknowledgments}
The authors would like to sincerely thank all anonymous reviewers and our shepherd for their comprehensive feedback, helpful suggestions, and the time they invested, which significantly improved the quality of this paper. This research/project is supported by the National Research Foundation, Singapore and DSO National Laboratories under its AI Singapore Programme (AISG Award No: AISG2-GC-2023-008). It is also supported by the National Research Foundation, Singapore, the Infocomm Media Development Authority under its Trust Tech Funding Initiative, the National Natural Science Foundation of China (NSFC) under Grants No. 62202340, and the Fundamental Research Funds for the Central Universities under No. 2042023kf0121. Any opinions, findings and conclusions or recommendations expressed in this material are those of the author(s) and do not reflect the views of National Research Foundation, Singapore and Infocomm Media Development Authority.






%



{\footnotesize \bibliography{main}}

\begin{thebibliography}{73}
\providecommand{\natexlab}[1]{#1}
\providecommand{\url}[1]{#1}
\csname url@samestyle\endcsname
\providecommand{\newblock}{\relax}
\providecommand{\bibinfo}[2]{#2}
\providecommand{\BIBentrySTDinterwordspacing}{\spaceskip=0pt\relax}
\providecommand{\BIBentryALTinterwordstretchfactor}{4}
\providecommand{\BIBentryALTinterwordspacing}{\spaceskip=\fontdimen2\font plus
\BIBentryALTinterwordstretchfactor\fontdimen3\font minus \fontdimen4\font\relax}
\providecommand{\BIBforeignlanguage}[2]{{%
\expandafter\ifx\csname l@#1\endcsname\relax
\typeout{** WARNING: IEEEtranN.bst: No hyphenation pattern has been}%
\typeout{** loaded for the language `#1'. Using the pattern for}%
\typeout{** the default language instead.}%
\else
\language=\csname l@#1\endcsname
\fi
#2}}
\providecommand{\BIBdecl}{\relax}
\BIBdecl

\bibitem[Rombach et~al.(2022)Rombach, Blattmann, Lorenz, Esser, and Ommer]{rombach2022stable}
R.~Rombach, A.~Blattmann, D.~Lorenz, P.~Esser, and B.~Ommer, ``High-resolution image synthesis with latent diffusion models,'' in \emph{CVPR}, 2022.

\bibitem[RunwayML(2024)]{runwayml2024stable}
RunwayML, ``Stable diffusion v1.5,'' \url{https://huggingface.co/runwayml/stable-diffusion-v1-5}, 2024.

\bibitem[Luo et~al.(2023{\natexlab{a}})Luo, Tan, Huang, Li, and Zhao]{luo2023latent}
S.~Luo, Y.~Tan, L.~Huang, J.~Li, and H.~Zhao, ``Latent consistency models: Synthesizing high-resolution images with few-step inference,'' \emph{arXiv preprint arXiv:2310.04378}, 2023.

\bibitem[Razzhigaev et~al.(2023)Razzhigaev, Shakhmatov, Maltseva, Arkhipkin, Pavlov, Ryabov, Kuts, Panchenko, Kuznetsov, and Dimitrov]{razzhigaev2023kandinsky}
A.~Razzhigaev, A.~Shakhmatov, A.~Maltseva, V.~Arkhipkin, I.~Pavlov, I.~Ryabov, A.~Kuts, A.~Panchenko, A.~Kuznetsov, and D.~Dimitrov, ``Kandinsky: An improved text-to-image synthesis with image prior and latent diffusion,'' in \emph{EMNLP}, 2023.

\bibitem[Ruiz et~al.(2023)Ruiz, Li, Jampani, Pritch, Rubinstein, and Aberman]{ruiz2023dreambooth}
N.~Ruiz, Y.~Li, V.~Jampani, Y.~Pritch, M.~Rubinstein, and K.~Aberman, ``Dreambooth: Fine tuning text-to-image diffusion models for subject-driven generation,'' in \emph{CVPR}, 2023.

\bibitem[Kumari et~al.(2023)Kumari, Zhang, Zhang, Shechtman, and Zhu]{kumari2023multi}
N.~Kumari, B.~Zhang, R.~Zhang, E.~Shechtman, and J.-Y. Zhu, ``Multi-concept customization of text-to-image diffusion,'' in \emph{CVPR}, 2023.

\bibitem[Han et~al.(2023)Han, Li, Zhang, Milanfar, Metaxas, and Yang]{han2023svdiff}
L.~Han, Y.~Li, H.~Zhang, P.~Milanfar, D.~Metaxas, and F.~Yang, ``Svdiff: Compact parameter space for diffusion fine-tuning,'' in \emph{ICCV}, 2023.

\bibitem[{Civitai}(2024)]{civitai2024}
\BIBentryALTinterwordspacing
{Civitai}, ``Civitai: The home of open-source generative ai,'' 2024. [Online]. Available: \url{https://civitai.com/}
\BIBentrySTDinterwordspacing

\bibitem[{Replicate}(2024)]{replicate2024}
{Replicate}, ``Replicate: Run machine learning models in the cloud,'' \url{https://replicate.com/}, 2024.

\bibitem[{Scenario}(2024)]{scenario2024}
{Scenario}, ``Scenario - ai-generated game assets,'' \url{https://www.scenario.com/}, 2024.

\bibitem[{Liblib AI}(2024)]{liblib_art}
\BIBentryALTinterwordspacing
{Liblib AI}, ``Liblib ai,'' 2024. [Online]. Available: \url{https://www.liblib.art/}
\BIBentrySTDinterwordspacing

\bibitem[Wang et~al.(2024)Wang, Chen, Lyu, Metaxas, and Ma]{wang2023diagnosis}
Z.~Wang, C.~Chen, L.~Lyu, D.~N. Metaxas, and S.~Ma, ``Diagnosis: Detecting unauthorized data usages in text-to-image diffusion models,'' in \emph{ICLR}, 2024.

\bibitem[Cui et~al.(2023)Cui, Ren, Xu, He, Liu, Sun, and Tang]{cui2023diffusionshield}
Y.~Cui, J.~Ren, H.~Xu, P.~He, H.~Liu, L.~Sun, and J.~Tang, ``Diffusionshield: A watermark for copyright protection against generative diffusion models,'' \emph{arXiv preprint arXiv:2306.04642}, 2023.

\bibitem[Shan et~al.(2023)Shan, Cryan, Wenger, Zheng, Hanocka, and Zhao]{shan2023glaze}
S.~Shan, J.~Cryan, E.~Wenger, H.~Zheng, R.~Hanocka, and B.~Y. Zhao, ``Glaze: Protecting artists from style mimicry by text-to-image models,'' in \emph{USENIX Security 23}, 2023.

\bibitem[Li et~al.(2023{\natexlab{a}})Li, Zhu, Yang, Jiang, Wei, and Xia]{li2023black}
Y.~Li, M.~Zhu, X.~Yang, Y.~Jiang, T.~Wei, and S.-T. Xia, ``Black-box dataset ownership verification via backdoor watermarking,'' \emph{IEEE TIFS}, 2023.

\bibitem[Sablayrolles et~al.(2020)Sablayrolles, Douze, Schmid, and J{\'e}gou]{sablayrolles2020radioactive}
A.~Sablayrolles, M.~Douze, C.~Schmid, and H.~J{\'e}gou, ``Radioactive data: tracing through training,'' in \emph{ICML}, 2020.

\bibitem[Guo et~al.(2024)Guo, Li, Wang, Xia, Huang, Liu, and Li]{guo2024domain}
J.~Guo, Y.~Li, L.~Wang, S.-T. Xia, H.~Huang, C.~Liu, and B.~Li, ``Domain watermark: Effective and harmless dataset copyright protection is closed at hand,'' \emph{NeurIPS}, 2024.

\bibitem[Tang et~al.(2023)Tang, Feng, Liu, Yang, and Hu]{tang2023did}
R.~Tang, Q.~Feng, N.~Liu, F.~Yang, and X.~Hu, ``Did you train on my dataset? towards public dataset protection with cleanlabel backdoor watermarking,'' \emph{ACM SIGKDD}, 2023.

\bibitem[Li et~al.(2022{\natexlab{a}})Li, Bai, Jiang, Yang, Xia, and Li]{li2022untargeted}
Y.~Li, Y.~Bai, Y.~Jiang, Y.~Yang, S.-T. Xia, and B.~Li, ``Untargeted backdoor watermark: Towards harmless and stealthy dataset copyright protection,'' \emph{NeurIPS}, 2022.

\bibitem[Yu et~al.(2021)Yu, Skripniuk, Abdelnabi, and Fritz]{yu2021artificial}
N.~Yu, V.~Skripniuk, S.~Abdelnabi, and M.~Fritz, ``Artificial fingerprinting for generative models: Rooting deepfake attribution in training data,'' in \emph{ICCV}, 2021.

\bibitem[Zhao et~al.(2023{\natexlab{a}})Zhao, Pang, Du, Yang, Cheung, and Lin]{zhao2023recipe}
Y.~Zhao, T.~Pang, C.~Du, X.~Yang, N.-M. Cheung, and M.~Lin, ``A recipe for watermarking diffusion models,'' \emph{arXiv preprint arXiv:2303.10137}, 2023.

\bibitem[Luo et~al.(2023{\natexlab{b}})Luo, Huang, Zhang, Qian, Li, and Zhang]{luo2023steal}
G.~Luo, J.~Huang, M.~Zhang, Z.~Qian, S.~Li, and X.~Zhang, ``Steal my artworks for fine-tuning? a watermarking framework for detecting art theft mimicry in text-to-image models,'' \emph{arXiv preprint arXiv:2311.13619}, 2023.

\bibitem[Maini et~al.(2021)Maini, Yaghini, and Papernot]{maini2020dataset}
P.~Maini, M.~Yaghini, and N.~Papernot, ``Dataset inference: Ownership resolution in machine learning,'' in \emph{ICLR}, 2021.

\bibitem[Somepalli et~al.(2023)Somepalli, Singla, Goldblum, Geiping, and Goldstein]{somepalli2023diffusion}
G.~Somepalli, V.~Singla, M.~Goldblum, J.~Geiping, and T.~Goldstein, ``Diffusion art or digital forgery? investigating data replication in diffusion models,'' in \emph{CVPR}, 2023.

\bibitem[AI(2024)]{stabilityai_stable_diffusion_2_1}
S.~AI, ``Stable diffusion 2-1,'' \url{https://huggingface.co/stabilityai/stable-diffusion-2-1}, 2024.

\bibitem[Van~Le et~al.(2023)Van~Le, Phung, Nguyen, Dao, Tran, and Tran]{van2023anti}
T.~Van~Le, H.~Phung, T.~H. Nguyen, Q.~Dao, N.~N. Tran, and A.~Tran, ``Anti-dreambooth: Protecting users from personalized text-to-image synthesis,'' in \emph{ICCV}, 2023.

\bibitem[Liu et~al.(2024)Liu, Fan, Dai, Chen, Zhou, and Sun]{liu2024metacloak}
Y.~Liu, C.~Fan, Y.~Dai, X.~Chen, P.~Zhou, and L.~Sun, ``Metacloak: Preventing unauthorized subject-driven text-to-image diffusion-based synthesis via meta-learning,'' in \emph{CVPR}, 2024.

\bibitem[Pang and Wang(2023)]{pang2023black}
Y.~Pang and T.~Wang, ``Black-box membership inference attacks against fine-tuned diffusion models,'' \emph{arXiv preprint arXiv:2312.08207}, 2023.

\bibitem[Duan et~al.(2023)Duan, Kong, Wang, Shi, and Xu]{duan2023diffusion}
J.~Duan, F.~Kong, S.~Wang, X.~Shi, and K.~Xu, ``Are diffusion models vulnerable to membership inference attacks?'' in \emph{ICML}, 2023.

\bibitem[Radford et~al.(2021)Radford, Kim, Hallacy, Ramesh, Goh, Agarwal, Sastry, Askell, Mishkin, Clark, et~al.]{radford2021learning}
A.~Radford, J.~W. Kim, C.~Hallacy, A.~Ramesh, G.~Goh, S.~Agarwal, G.~Sastry, A.~Askell, P.~Mishkin, J.~Clark \emph{et~al.}, ``Learning transferable visual models from natural language supervision,'' in \emph{ICML}, 2021.

\bibitem[Oquab et~al.(2024)Oquab, Darcet, Moutakanni, Vo, Szafraniec, Khalidov, Fernandez, HAZIZA, Massa, El-Nouby, Assran, Ballas, Galuba, Howes, Huang, Li, Misra, Rabbat, Sharma, Synnaeve, Xu, Jegou, Mairal, Labatut, Joulin, and Bojanowski]{oquab2024dinov}
M.~Oquab, T.~Darcet, T.~Moutakanni, H.~V. Vo, M.~Szafraniec, V.~Khalidov, P.~Fernandez, D.~HAZIZA, F.~Massa, A.~El-Nouby, M.~Assran, N.~Ballas, W.~Galuba, R.~Howes, P.-Y. Huang, S.-W. Li, I.~Misra, M.~Rabbat, V.~Sharma, G.~Synnaeve, H.~Xu, H.~Jegou, J.~Mairal, P.~Labatut, A.~Joulin, and P.~Bojanowski, ``{DINO}v2: Learning robust visual features without supervision,'' \emph{TMLR}, 2024.

\bibitem[Caron et~al.(2021)Caron, Touvron, Misra, J{\'e}gou, Mairal, Bojanowski, and Joulin]{caron2021emerging}
M.~Caron, H.~Touvron, I.~Misra, H.~J{\'e}gou, J.~Mairal, P.~Bojanowski, and A.~Joulin, ``Emerging properties in self-supervised vision transformers,'' in \emph{ICCV}, 2021.

\bibitem[Li et~al.(2022{\natexlab{b}})Li, Li, Xiong, and Hoi]{li2023blip}
J.~Li, D.~Li, C.~Xiong, and S.~Hoi, ``Blip: Bootstrapping language-image pre-training for unified vision-language understanding and generation,'' in \emph{ICML}, 2022.

\bibitem[Kaggle(2019)]{pratama2019pokemon}
Kaggle, ``Pokemon images dataset,'' \url{https://www.kaggle.com/datasets/kvpratama/pokemon-images-dataset}, 2019.

\bibitem[Karras et~al.(2018)Karras, Aila, Laine, and Lehtinen]{karras2018progressive}
T.~Karras, T.~Aila, S.~Laine, and J.~Lehtinen, ``Progressive growing of {GAN}s for improved quality, stability, and variation,'' in \emph{ICLR}, 2018.

\bibitem[Liao et~al.(2022)Liao, Li, Liu, and Keutzer]{liao2022artbench}
P.~Liao, X.~Li, X.~Liu, and K.~Keutzer, ``The artbench dataset: Benchmarking generative models with artworks,'' \emph{arXiv preprint arXiv:2206.11404}, 2022.

\bibitem[Kaggle(2024)]{arnaud58_landscape_pictures}
Kaggle, ``Landscape pictures,'' \url{https://www.kaggle.com/datasets/arnaud58/landscape-pictures/data}, 2024.

\bibitem[Heusel et~al.(2017)Heusel, Ramsauer, Unterthiner, Nessler, and Hochreiter]{heusel2017gans}
M.~Heusel, H.~Ramsauer, T.~Unterthiner, B.~Nessler, and S.~Hochreiter, ``Gans trained by a two time-scale update rule converge to a local nash equilibrium,'' \emph{NeurIPS}, 2017.

\bibitem[Welch(1947)]{welch1947generalization}
B.~L. Welch, ``The generalization of ‘student's’problem when several different population varlances are involved,'' \emph{Biometrika}, 1947.

\bibitem[Ilyas et~al.(2019)Ilyas, Santurkar, Tsipras, Engstrom, Tran, and Madry]{ilyas2019adversarial}
A.~Ilyas, S.~Santurkar, D.~Tsipras, L.~Engstrom, B.~Tran, and A.~Madry, ``Adversarial examples are not bugs, they are features,'' \emph{NeurIPS}, 2019.

\bibitem[Tsipras et~al.(2018)Tsipras, Santurkar, Engstrom, Turner, and Madry]{tsipras2018robustness}
D.~Tsipras, S.~Santurkar, L.~Engstrom, A.~Turner, and A.~Madry, ``Robustness may be at odds with accuracy,'' in \emph{ICLR}, 2018.

\bibitem[Kwon et~al.(2023)Kwon, Jeong, and Uh]{kwon2022diffusion}
M.~Kwon, J.~Jeong, and Y.~Uh, ``Diffusion models already have a semantic latent space,'' in \emph{ICLR}, 2023.

\bibitem[Li et~al.(2023{\natexlab{b}})Li, Prabhudesai, Duggal, Brown, and Pathak]{li2023your}
A.~C. Li, M.~Prabhudesai, S.~Duggal, E.~Brown, and D.~Pathak, ``Your diffusion model is secretly a zero-shot classifier,'' in \emph{ICCV}, 2023.

\bibitem[Luo et~al.(2001)Luo, Cui, and Rigg]{luo2001development}
M.~R. Luo, G.~Cui, and B.~Rigg, ``The development of the cie 2000 colour-difference formula: Ciede2000,'' \emph{Color Research \& Application}, 2001.

\bibitem[Sharma et~al.(2005)Sharma, Wu, and Dalal]{sharma2005ciede2000}
G.~Sharma, W.~Wu, and E.~N. Dalal, ``The ciede2000 color-difference formula: Implementation notes, supplementary test data, and mathematical observations,'' \emph{Color Research \& Application}, 2005.

\bibitem[Ruff et~al.(2018)Ruff, Vandermeulen, Goernitz, Deecke, Siddiqui, Binder, M{\"u}ller, and Kloft]{ruff2018deep}
L.~Ruff, R.~Vandermeulen, N.~Goernitz, L.~Deecke, S.~A. Siddiqui, A.~Binder, E.~M{\"u}ller, and M.~Kloft, ``Deep one-class classification,'' in \emph{ICML}, 2018.

\bibitem[Forsythe et~al.(1977)]{forsythe1977computer}
G.~E. Forsythe \emph{et~al.}, \emph{Computer methods for mathematical computations}.\hskip 1em plus 0.5em minus 0.4em\relax Prentice-hall, 1977.

\bibitem[Fernandez et~al.(2023)Fernandez, Couairon, J{\'e}gou, Douze, and Furon]{fernandez2023stable}
P.~Fernandez, G.~Couairon, H.~J{\'e}gou, M.~Douze, and T.~Furon, ``The stable signature: Rooting watermarks in latent diffusion models,'' in \emph{ICCV}, 2023.

\bibitem[Massey~Jr(1951)]{massey1951kolmogorov}
F.~J. Massey~Jr, ``The kolmogorov-smirnov test for goodness of fit,'' \emph{Journal of the American statistical Association}, 1951.

\bibitem[Marsaglia et~al.(2003)Marsaglia, Tsang, and Wang]{marsaglia2003evaluating}
G.~Marsaglia, W.~W. Tsang, and J.~Wang, ``Evaluating kolmogorov's distribution,'' \emph{Journal of statistical software}, 2003.

\bibitem[Nichol and Schulman(2018)]{nichol2018reptile}
A.~Nichol and J.~Schulman, ``Reptile: a scalable metalearning algorithm,'' \emph{arXiv preprint arXiv:1803.02999}, 2018.

\bibitem[Gu et~al.(2022)Gu, Chen, Bao, Wen, Zhang, Chen, Yuan, and Guo]{gu2022vector}
S.~Gu, D.~Chen, J.~Bao, F.~Wen, B.~Zhang, D.~Chen, L.~Yuan, and B.~Guo, ``Vector quantized diffusion model for text-to-image synthesis,'' in \emph{CVPR}, 2022.

\bibitem[Saleh and Elgammal(2015)]{saleh2015large}
B.~Saleh and A.~Elgammal, ``Large-scale classification of fine-art paintings: Learning the right metric on the right feature,'' \emph{arXiv preprint arXiv:1505.00855}, 2015.

\bibitem[Hu et~al.(2022)Hu, yelong shen, Wallis, Allen-Zhu, Li, Wang, Wang, and Chen]{hu2022lora}
E.~J. Hu, yelong shen, P.~Wallis, Z.~Allen-Zhu, Y.~Li, S.~Wang, L.~Wang, and W.~Chen, ``Lo{RA}: Low-rank adaptation of large language models,'' in \emph{ICLR}, 2022.

\bibitem[Hore and Ziou(2010)]{hore2010image}
A.~Hore and D.~Ziou, ``Image quality metrics: Psnr vs. ssim,'' in \emph{ICPR}, 2010.

\bibitem[Wang et~al.(2004)Wang, Bovik, Sheikh, and Simoncelli]{wang2004image}
Z.~Wang, A.~C. Bovik, H.~R. Sheikh, and E.~P. Simoncelli, ``Image quality assessment: from error visibility to structural similarity,'' \emph{IEEE TIP}, 2004.

\bibitem[Zhang et~al.(2018)Zhang, Isola, Efros, Shechtman, and Wang]{zhang2018unreasonable}
R.~Zhang, P.~Isola, A.~A. Efros, E.~Shechtman, and O.~Wang, ``The unreasonable effectiveness of deep features as a perceptual metric,'' in \emph{CVPR}, 2018.

\bibitem[Jia et~al.(2021)Jia, Choquette-Choo, Chandrasekaran, and Papernot]{jia2021entangled}
H.~Jia, C.~A. Choquette-Choo, V.~Chandrasekaran, and N.~Papernot, ``Entangled watermarks as a defense against model extraction,'' in \emph{USENIX Security 21}, 2021.

\bibitem[Liu et~al.(2023)Liu, Li, Wu, and Lee]{liu2023llava}
H.~Liu, C.~Li, Q.~Wu, and Y.~J. Lee, ``Visual instruction tuning,'' in \emph{NeurIPS}, 2023.

\bibitem[Chen et~al.(2023)Chen, Wang, Beyer, Kolesnikov, Wu, Voigtlaender, Mustafa, Goodman, Alabdulmohsin, Padlewski, et~al.]{chen2023pali}
X.~Chen, X.~Wang, L.~Beyer, A.~Kolesnikov, J.~Wu, P.~Voigtlaender, B.~Mustafa, S.~Goodman, I.~Alabdulmohsin, P.~Padlewski \emph{et~al.}, ``Pali-3 vision language models: Smaller, faster, stronger,'' \emph{arXiv preprint arXiv:2310.09199}, 2023.

\bibitem[Reiss et~al.(2021)Reiss, Cohen, Bergman, and Hoshen]{reiss2021panda}
T.~Reiss, N.~Cohen, L.~Bergman, and Y.~Hoshen, ``Panda: Adapting pretrained features for anomaly detection and segmentation,'' in \emph{CVPR}, 2021.

\bibitem[Cheng et~al.(2020)Cheng, Sun, Takeuchi, and Katto]{cheng2020learned}
Z.~Cheng, H.~Sun, M.~Takeuchi, and J.~Katto, ``Learned image compression with discretized gaussian mixture likelihoods and attention modules,'' in \emph{CVPR}, 2020.

\bibitem[Zhao et~al.(2023{\natexlab{b}})Zhao, Zhang, Su, Vasan, Grishchenko, Kruegel, Vigna, Wang, and Li]{xu2023invisible}
X.~Zhao, K.~Zhang, Z.~Su, S.~Vasan, I.~Grishchenko, C.~Kruegel, G.~Vigna, Y.-X. Wang, and L.~Li, ``Invisible image watermarks are provably removable using generative ai,'' \emph{arXiv preprint arXiv: 2306.01953}, 2023.

\bibitem[Jiang et~al.(2023)Jiang, Zhang, and Gong]{JiangZG23}
Z.~Jiang, J.~Zhang, and N.~Z. Gong, ``Evading watermark based detection of ai-generated content,'' in \emph{CCS}, 2023.

\bibitem[Zeng et~al.(2023)Zeng, Pan, Just, Lyu, Qiu, and Jia]{zeng2023narcissus}
Y.~Zeng, M.~Pan, H.~A. Just, L.~Lyu, M.~Qiu, and R.~Jia, ``Narcissus: A practical clean-label backdoor attack with limited information,'' in \emph{CCS}, 2023.

\bibitem[Li et~al.(2021)Li, Lyu, Koren, Lyu, Li, and Ma]{li2021anti}
Y.~Li, X.~Lyu, N.~Koren, L.~Lyu, B.~Li, and X.~Ma, ``Anti-backdoor learning: Training clean models on poisoned data,'' \emph{NeurIPS}, 2021.

\bibitem[Phimsiri(2020)]{phimsiri2020animechibi}
\BIBentryALTinterwordspacing
H.~Phimsiri, ``Anime chibi datasets,'' 2020, accessed: 2024-10-07. [Online]. Available: \url{https://www.kaggle.com/datasets/hirunkulphimsiri/anime-chibi-datasets}
\BIBentrySTDinterwordspacing

\bibitem[Ronneberger et~al.(2015)Ronneberger, Fischer, and Brox]{unet}
O.~Ronneberger, P.~Fischer, and T.~Brox, ``U-net: Convolutional networks for biomedical image segmentation,'' in \emph{MICCAI}, 2015.

\bibitem[Zirpoli(2023)]{zirpoli2023generative}
C.~T. Zirpoli, ``Generative artificial intelligence and copyright law,'' 2023.

\bibitem[Zhu et~al.(2018)Zhu, Kaplan, Johnson, and Fei-Fei]{Zhu_2018_ECCV}
J.~Zhu, R.~Kaplan, J.~Johnson, and L.~Fei-Fei, ``Hidden: Hiding data with deep networks,'' in \emph{ECCV}, 2018.

\bibitem[Nguyen et~al.(2015)Nguyen, Yosinski, and Clune]{Nguyen_2015_CVPR}
A.~Nguyen, J.~Yosinski, and J.~Clune, ``Deep neural networks are easily fooled: High confidence predictions for unrecognizable images,'' in \emph{CVPR}, 2015.

\bibitem[Hampton and Bailey(2020)]{Hampton2020}
\BIBentryALTinterwordspacing
S.~D. Hampton and A.~J. Bailey, ``Intellectual property case filing trends over the last decade,'' 2020, accessed: 2024-10-12. [Online]. Available: \url{https://www.hamptonip.com/articles/post/intellectual-property-case-filing-trends-over-the-last-decade/}
\BIBentrySTDinterwordspacing

\bibitem[Alemohammad et~al.(2024)Alemohammad, Casco-Rodriguez, Luzi, Humayun, Babaei, LeJeune, Siahkoohi, and Baraniuk]{alemohammad2024selfconsuming}
S.~Alemohammad, J.~Casco-Rodriguez, L.~Luzi, A.~I. Humayun, H.~Babaei, D.~LeJeune, A.~Siahkoohi, and R.~Baraniuk, ``Self-consuming generative models go {MAD},'' in \emph{ICLR}, 2024.

\end{thebibliography}
\bibliographystyle{IEEEtranN}


\appendices
\section{FPR Empirical Check}
\label{app:fpr}
\vspace{-0.5em}
Recall that the significance level $\alpha$ in Eq. (\ref{eq:kstest}) controls the probability of the test for making Type-I error \ie rejecting the null hypothesis while it is actually true (false positives). Additionally, we are interested in whether the choice of benign distribution (\ie $G$ in Eq. (\ref{eq:kstest})) largely impacts the FPR of {\sc Siren}. As such, we examine whether the FPR controlled by $\alpha$ aligns with empirical observations with different choices of benign samples.

\begin{figure}[t]
  \centering
    \begin{subfigure}{0.156\textwidth}
        \centering
        \includegraphics[width=\linewidth]{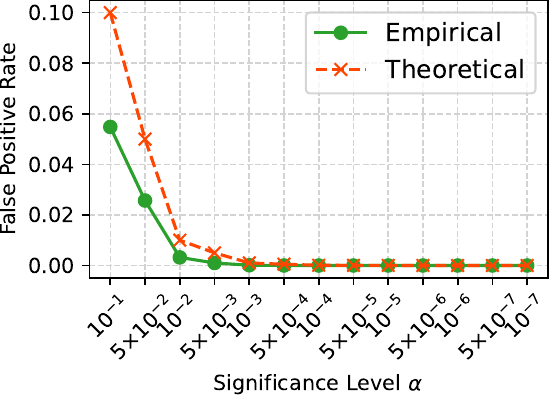}
        \caption{Original SD}
    \end{subfigure}
    \hfill
    \begin{subfigure}{0.156\textwidth}
        \centering
        \includegraphics[width=\linewidth]{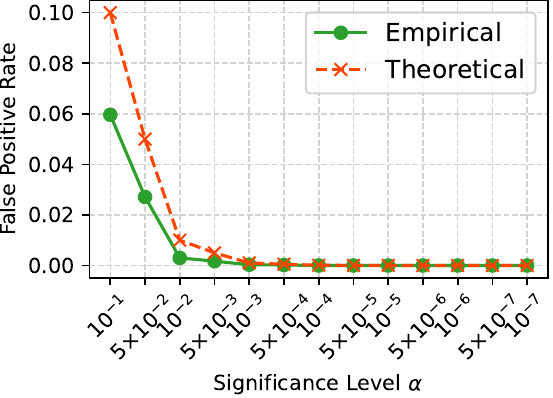}
        \caption{Fine-tuned SD}
    \end{subfigure}    
    \begin{subfigure}{0.156\textwidth}
        \centering
        \includegraphics[width=\linewidth]{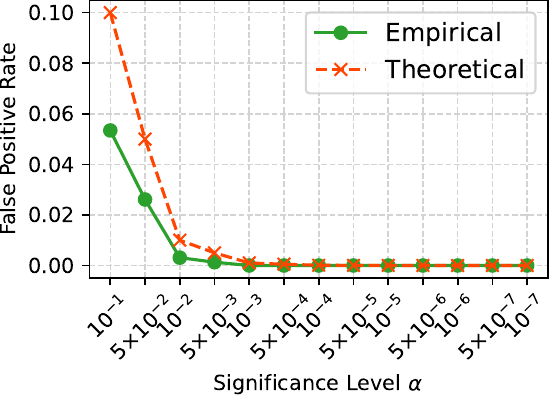}
        \caption{Kandinsky}
    \end{subfigure}    
  \caption{\small FPR empirical check. Original SD, Fine-tuned SD, and Kandinsky refers to using the original Stable Diffusion v1.5, the Stable Diffusion v1.5 fine-tuned on the uncoated version of Pokemon, and the Kandinsky 2.2 to serve as the benign model.}
  \label{fig:fpr-check}
  \vspace{-0.5em}
\end{figure}

Figure \ref{fig:fpr-check} shows that the empirical FPR of our method closely follows the controlled one (the ``theoretical'' line) and is mostly lower than $\alpha$, this is possibly because K-S test is conservative when the sample size is small. Besides, we can see that \sys is not sensitive to the choice of benign model. This is because our extractor is only responsive to the injected coating (instead of image content), so the choice of the benign model does not have a huge impact on the test result regardless of the negative sample choices.

\section{Additional Experiments}
\label{app:exp}
\vspace{-0.5em}
In this section, we present more additional experiments to further analyze \sys and discuss other potential scenarios that \sys may encounter in the real world.

\begin{figure*}[t]
  \begin{subfigure}[b]{0.235\textwidth}
    \includegraphics[width=\textwidth]{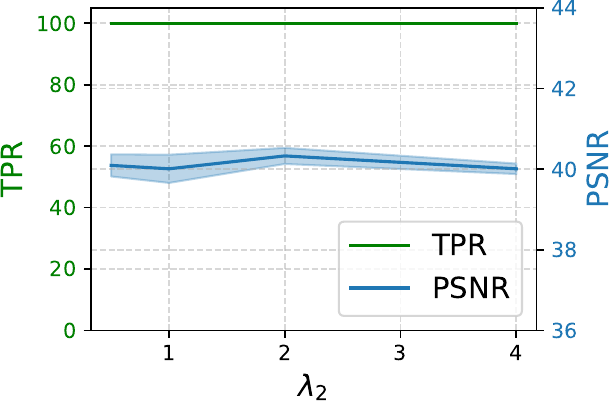}
    \caption{$\lambda_1=0.5$}
  \end{subfigure}
  \hfill
  \begin{subfigure}[b]{0.235\textwidth}
    \includegraphics[width=\textwidth]{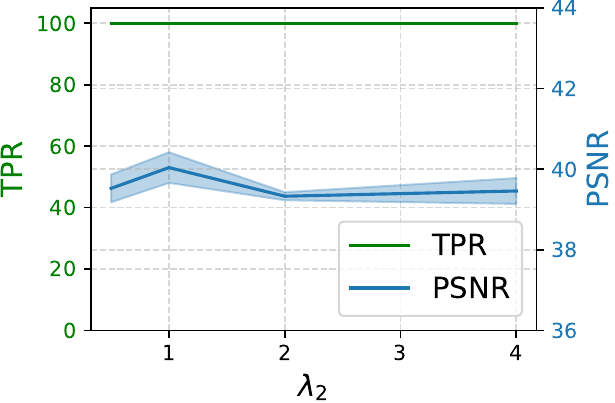}
    \caption{$\lambda_1=1$}
  \end{subfigure}
  \hfill
  \begin{subfigure}[b]{0.235\textwidth}
    \includegraphics[width=\textwidth]{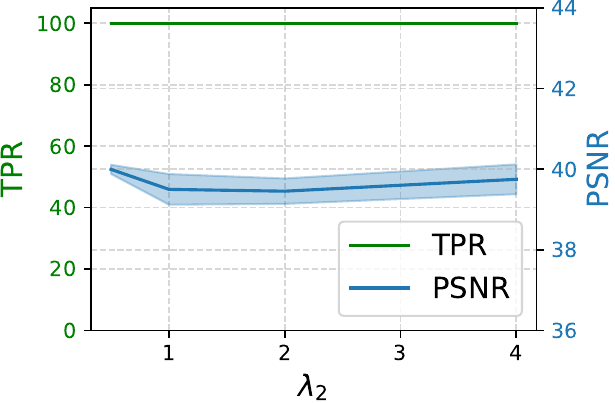}
    \caption{$\lambda_1=2$}
  \end{subfigure}
  \hfill
  \begin{subfigure}[b]{0.235\textwidth}
    \includegraphics[width=\textwidth]{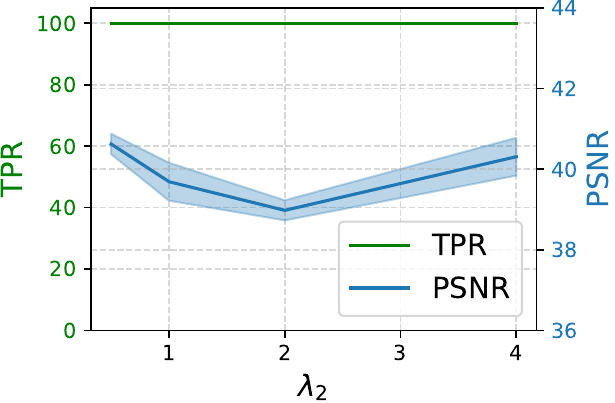}
    \caption{$\lambda_1=4$}
  \end{subfigure}
  \caption{{Ablation study on weighting parameters.} We repeat each experiment three times.}
  \vspace{-1em}
  \label{fig:ablation}
\end{figure*}

\vspace{0.3em}
\noindent\textbf{The mimicries undergo image transformations before verification.} Following previous works \citep{wang2023diagnosis,yu2021artificial}, we evaluate the robustness of \sys when the mimicries are transformed before verification. We consider a comprehensive list of 13 types of common image transformations that may happen in practice. As shown in Table \ref{tab:robustness-datasets}, the robustness of \sys remains surprisingly high for these transformations. A noteworthy case is Crop $0.1$, where \sys remains a $100\% \text{TPR}$  when only the center $10\%$ of the image is left. The robustness of \sys can be mainly attributed to the EoT layer, which has been widely verified to enhance the robustness against image transformations.

\begin{table}[t]
    \centering
    \caption{\small Evaluation on robustness of \sys against common image transformations. Cont. refers to contrast, Sup. Res. refers to first downscaling the image to 0.7 and then upscaling it via a super-resolution model, and Rand. Comb. refers to a random combination of all transformations. $\alpha$ is set to $10^{-9}$.}
    \label{tab:robustness-datasets}
    \footnotesize
    \setlength{\tabcolsep}{4pt}
    \resizebox{0.96\linewidth}{!}{
        \begin{tabular}{ll|ll|ll}
            \toprule
            \multicolumn{6}{c}{\textbf{(a) Pokemon Dataset}} \\
            \midrule
            \bf{Attack} & \bf{TPR $\uparrow$} & Noise $0.2$ & $100\%$ & Sharpness $2.0$ & $100\%$ \\
            None & $100\%$ & Bright. $1.5$ & $100\%$ & Blur ($k$=7) & $100\%$ \\
            Crop $0.1$ & $100\%$ & Cont. $2.0$ & $100\%$ & Sup. Res. $0.7$ & $100\%$ \\
            Hue  & $100\%$ & Quantize 8bit & $100\%$ & Text Overlay & $100\%$ \\
            JPEG $30$ & $100\%$ & Sat. $2.0$ & $100\%$ & Rand. Comb. & $100\%$ \\
            \midrule
            \multicolumn{6}{c}{\textbf{(b) CelebA-HQ Dataset}} \\
            \midrule
            \bf{Attack} & \bf{TPR $\uparrow$} & Noise $0.2$ & $100\%$ & Sharpness $2.0$ & $100\%$ \\
            None & $100\%$ & Bright. $1.5$ & $99.9\%$ & Blur ($k$=7) & $100\%$ \\
            Crop $0.1$ & $100\%$ & Cont. $2.0$ & $100\%$ & Sup. Res. $0.7$ & $100\%$ \\
            Hue  & $100\%$ & Quantize 8bit & $100\%$ & Text Overlay & $100\%$ \\
            JPEG $30$ & $100\%$ & Sat. $2.0$ & $100\%$ & Rand. Comb. & $100\%$ \\
            \bottomrule
        \end{tabular}
    }
    \vspace{-0.5em}
\end{table}

\begin{figure}[t]
    \centering
    \includegraphics[width=0.7\linewidth]{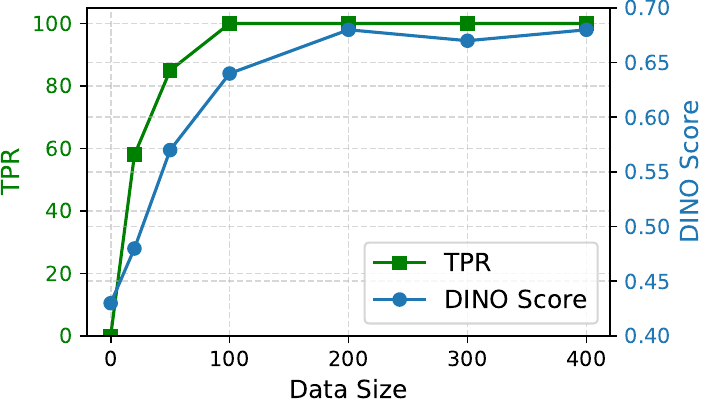}
    \caption{\small The effectiveness of \sys under different data sizes.}
    \label{fig:datasize}
    \vspace{-1.5em}
\end{figure}

\vspace{0.3em}
\noindent\textbf{{\sc Siren}'s effectiveness under different data sizes.} We discover the {\sc Siren}'s effectiveness under different data sizes. Specifically, we coat the Pokemon dataset and select a subset from it with different data sizes. Then we use the selected subset to train a personalized Stable Diffusion v1.5 model and measure  (1) whether personalization training is successful, and (2) whether \sys can effectively detect the coating from the generated images. As shown in Figure \ref{fig:datasize}, we observe that \sys is highly effective as long as the data size is sufficient for successful personalization learning.

\vspace{0.3em}
\noindent\textbf{The model undergoes further modifications.} We also evaluate whether the effectiveness of \sys degrades when the model undergoes further modifications after training. In detail, we first train a personalized model on a coated Pokemon dataset and tried (1) fine-tune this model further on a similar dataset \citep{phimsiri2020animechibi} using the DreamBooth method; and (2) quantize the model to 16-bit; We find \sys is still effective, retaining a TPR of 100\% in these cases.

\begin{table}[t]
    \centering
    \caption{\small Effectiveness with different generation settings on Pokemon dataset. In the Prompts column, ``class-based'' refers to using ``an image of [class]'' as prompts, ``validation prompt'' refers to using the prompts from the Pokemon validation set as prompts, and ``LLM-generated'' refers to providing the training set prompts (we randomly select 50 prompts) to GPT-4 and ask it to generate diverse prompts with similar contents. $\alpha$ is set to $10^{-9}$.} 
    \label{tab:generation_params}
    \footnotesize
    \vspace*{-0.2em}
    \setlength{\tabcolsep}{6pt}
    \resizebox{0.96\linewidth}{!}{
        \begin{tabular}{cccccc}
            \toprule
            \multicolumn{2}{c}{\bf{Configurations}} & \bf{TPR $\uparrow$} & \multicolumn{2}{c}{\bf{Configurations}} & \bf{TPR $\uparrow$} \\
            \midrule
            \multirow{3}{*}{Prompts} & Class-based & 100\% & \multirow{3}{*}{Sampling Steps} & 15 &100\%\\
                                     & Validation prompt & 100\%& & 25 &100\%\\
                                     & LLM-generated & 100\%& & 35 &100\%\\
            \midrule
            \multirow{4}{*}{Sampler} & DDPM & 100\% & \multirow{2}{*}{CFG Scale} & 5.0 & 100\%\\
            & DDIM & 100\% & & 7.5 & 100\%\\
            & Euler & 100\% & \multirow{2}{*}{Clip Skip} & 4 & 100\%\\
            & DPM++ & 100\% & & 6 & 100\%\\
            \bottomrule
        \end{tabular}
    }
    \vspace{-1.5em}
\end{table}

\vspace{0.3em}
\noindent\textbf{Different generation hyperparameters and prompts.} We investigate whether \sys remains effective under different generation hyperparameters and generation prompts. As shown in Table \ref{tab:generation_params}, our \sys has high effectiveness with different generation settings and prompts.

\vspace{0.3em}
\noindent\textbf{Hyperparameter analysis.} The relative strength of the losses is controlled by the weighting parameters $\lambda_1$ and $\lambda_2$, which are the key hyperparameters of our {\sc Siren}. Note that we set them to both 1 (meaning equal initial scales) to show equal importance. As illustrated in Figure \ref{fig:ablation}, SIREN is not sensitive to the choice of weighting parameters. Thus, we empirically set both of them to 1 in our experiments.

\section{Omitted Algorithm \& Experimental Details}
\label{app:details}
\vspace{-0.5em}
\noindent\textbf{Omitted details for our generator/extractor design.} Our coating generator and extractor network architecture follows the design in HiDDeN \citep{Zhu_2018_ECCV}, the only change is we discard the bit string in the generator input, and the final layer of the decoder is abandoned (we only need the feature space). Such architecture is very lightweight, it takes less than 2 MB to store a generator/extractor pair. We use the implementation from \citet{fernandez2023stable}, which adds an additional just noticeable difference layer for better perceptual quality. As raw perceptual loss may be unstable, we added an MSE image loss with a weighting parameter of 1 and included this layer during training and found that it helps stabilize. We also follow \citep{fernandez2023stable} to include an Expectation over Transformation (EoT) layer into the extractor, which is widely used to enhance robustness against real-world distortions. It simulates such distortions through a differentiable layer in train time before sending the image to the extractor, so the extractor will learn to be robust against them.

\setlength{\textfloatsep}{12pt} 
\begin{algorithm}[t]
\small
\caption{Meta learning with Reptile}\label{alg:reptile}
\begin{algorithmic}[1]
\Statex \textbf{Input}: Coating generator $\mathcal{G}$ and extractor $\Phi$, inner loop learning rate $\alpha$, $\beta$, meta-learning rate $\gamma$, $\xi$, training iterations $N$, 
number of inner loop iterations $K$
\For{$i = 1, \cdots, N$}
    \While{not all batches have been sampled}
    \State $\mathcal{G}^*_0,{\Phi}^*_{0} = \text{Clone}(\mathcal{G}^*,{\Phi}^*)$
        \State Sample a batch $\mathcal{D}_p$ from the training set
        \For{$k = 1, \cdots, K$}
            \State Calculate $\mathcal{L}_{\text{overall}}$ with $\mathcal{G}^*_{k-1}$ and ${\Phi}^*_{k-1}$ via Eq. (\ref{eq:overall})
            \State $\mathcal{G}^*_k \leftarrow \mathcal{G}^*_{k-1}-\alpha \nabla_{\mathcal{G}^*_{k-1}} \mathcal{L}_{\text{overall}}$
            \State ${\Phi}^*_{k} \leftarrow {\Phi}^*_{k-1}-\beta \nabla_{{\Phi}^*_{k-1}} \mathcal{L}_{\text{overall}}$
        \EndFor
    \EndWhile
    \State    $\mathcal{G}^* \leftarrow \mathcal{G}^* - \gamma (\mathcal{G}^*-\mathcal{G}^*_K)$
    \State $\Phi^* \leftarrow \Phi^* - \xi (\Phi^*-\Phi^*_K)$
\EndFor
\end{algorithmic}
\end{algorithm}

\vspace{0.3em}
\noindent\textbf{Omitted details for meta-learning.} The proxy dataset is MS-COCO, consisting of about 120,000 daily-life images and corresponding text descriptions. The omitted algorithm details of our meta-learning are presented in Algorithm \ref{alg:reptile}. The default learning rate $\alpha$ and $\beta$ is 1e-3, and the default meta learning rate $\xi$ and $\gamma$ is 1e-2. Intuitively, in each iteration, meta-learning samples a batch of text-image pairs from MS-COCO and uses them to fine-tune the meta-model. Then, it uses the parameter differences as the meta-gradient to update the meta-model. Through this process, the meta-model learns a set of initial weights that can quickly adapt to new datasets using a few fine-tuning steps. Note that for extremely small data sizes (\eg less than 10), it is still challenging to obtain satisfying models even with the help of meta-learning. To remedy this, we use the Stable Diffusion v1.5 model to generate 100 samples using BLIP-generated prompts of the data to supplement the training set.

\vspace{0.3em}
\noindent\textbf{Details on datasets and baseline configurations.} For Pokemon (833 high-quality Pokemon images) and Dog (5 high-quality images of a specific dog), we use the full training set. Since fine-tuning usually does not require much data, for CelebA-HQ (human facial images), Artbench (artworks from different classical artists), and Landscape (real-world landscape images), we randomly select 1,000 images for training. This practice aligns with \citep{wang2023diagnosis}. For Wikiart, we construct a subset consisting of 10 artists from 5 different art styles, with 25 to 40 artworks for each artist, and the reported TPR is averaged across all artists. 

For watermark-based baselines, the TPR is calculated by determining whether the total matched bits (for all samples) exceed a certain threshold $t$. FPR (or p-value) is regarded as the chance to achieve or exceed this threshold and can be obtained from the CDF of the binomial distribution.  It is calculated as $\sum_{i=t}^{n \times k}\binom{n \times k}{i}0.5^{n \times k}$, where $n$ is the sample size (30 in this paper) and $k$ is the bit length (32 for both baselines). For \citep{yu2021artificial}, the watermark encoder-decoder is trained following their original code implementation on $512\times 512$ MS-COCO dataset. For \citep{luo2023steal} which used four watermarking schemes, we choose DCT-DWT-SVD, which has the highest ``best bits'' as reported in \citep{luo2023steal}. We coat all images with the same watermark string before training.

\begin{figure}[t]
    \centering
    \includegraphics[width=\linewidth]{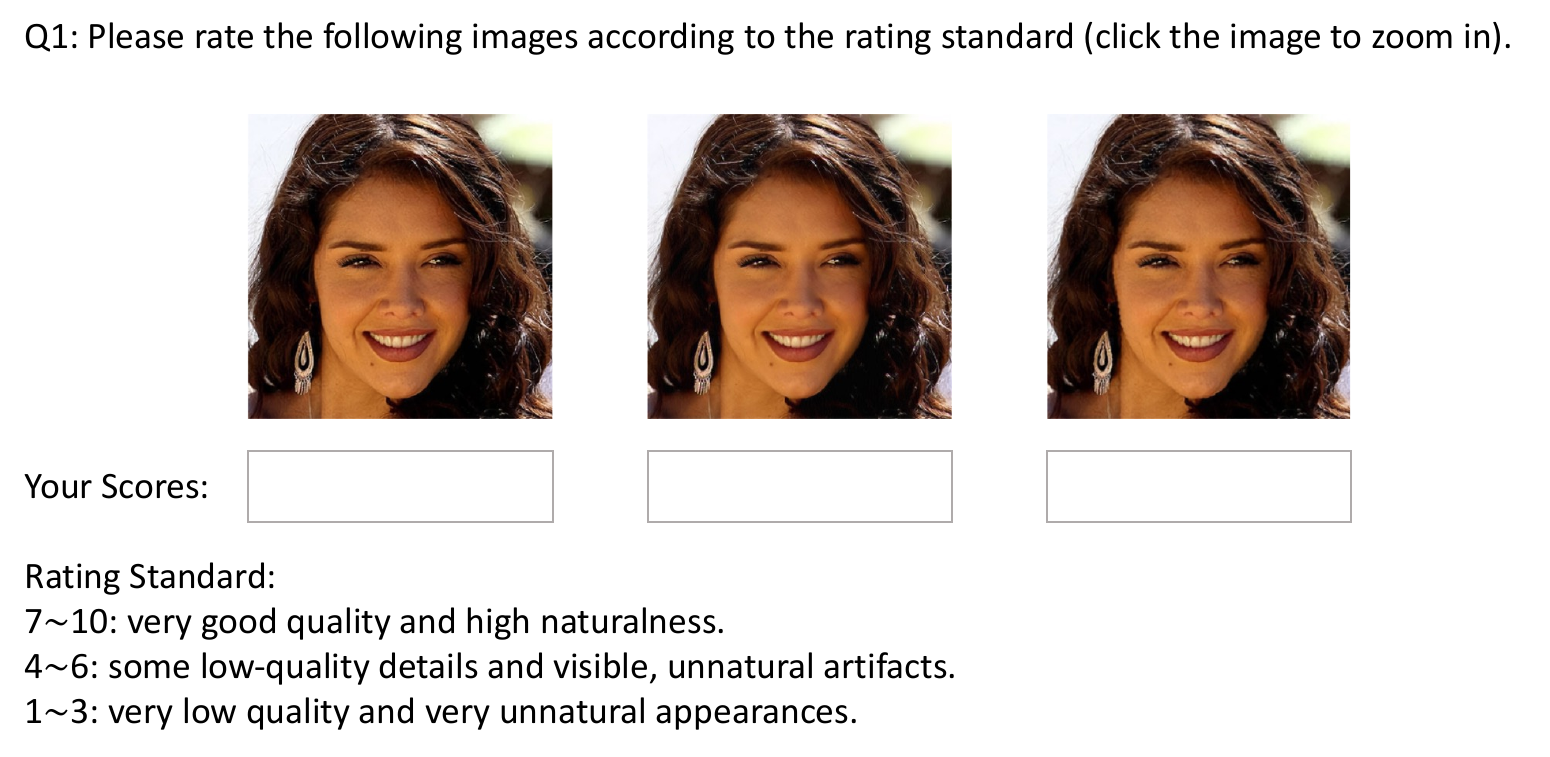}
    \caption{A sample question from our survey.}
    \label{fig:sample}
    \vspace{-1.5em}
\end{figure}

For backdoor-based baseline DIAGNOSIS \citep{wang2023diagnosis}, the hypothesis testing procedure follows its original paper. This test is supported by the theoretical analysis in \citep{li2023black}. In Table \ref{tab:diagnosis-lim}, we use the ``trigger-conditional'' (warping strength=1.0, text trigger ``tq'' in training prompts, and 20\% coating rate) method in \citep{wang2023diagnosis}. In the main experiments, for a fair comparison, we employ the ``unconditional'' (warping strength=2.0, training prompts same as other methods and 100\% coating rate) setting for DIAGNOSIS. The sample size for all tests is set to $30$, aligned with our method and other baselines.

\vspace{0.3em}
\noindent\textbf{More details about our human preference study.} We use several methods to improve the reliability of our human preference study. To ensure our answers are valid, we have thorough instructions for each tester before taking the survey. We explained the basics of \sys and the baselines and sought for their confirmation of understanding the task. We also manually checked each survey answer to be complete. To mitigate biases, we ensured diversity in gender and age among the participants. For all questions, the images are listed in random order. To ensure consistency, we have manually checked all outlier answers and asked the participants additional questions on the basics of \sys to confirm they fully understand the tasks. We provide a sample question from our survey for human evaluation in Figure \ref{fig:sample}. For each question, three images (original, coated by \sys, coated by DIAGNOSIS) are shown in a row, and the rating standard is also shown below each figure. The participants can click the image to zoom in and view the details. 

\vspace{0.3em}
\noindent\textbf{Detailed design of ABL.} ABL \citep{li2021anti} is a training-time backdoor defense that leverages the difference of loss scales between poisoned and benign training samples. In this paper, we adopt it to diffusion models as an adaptive countermeasure against {\sc Siren}. ABL is divided into two stages: for the first stage, it exploits the observation that the coating (trigger) feature can possibly be learned faster than other features. Therefore, it utilizes the loss drop characteristics to filter the likely coated samples. The filer rate (isolation rate) of ABL is set to 5\% in this stage. Since the loss value for diffusion models not only depends on the sample but also the timestep $t$, we calculate the expected loss as the averaged loss across $[100, 400, 700]$ time steps. The loss threshold $\gamma$ is set as the 5th percentile of expected loss across all samples on the Stable Diffusion v1.5 base model. In the second stage, the attack maximizes the loss of filtered (poisoned/coated) samples, \ie unlearning them. We directly reverse the sign of diffusion MSE loss to achieve this. It would help the model identify potential coating patterns and forget them.

\section{Discussions}
\label{app:discussion}
\vspace{-0.5em}
\noindent\textbf{Can we use DINO score to identify data infringement?} Notably, the CLIP score \citep{radford2021learning} and DINO score \citep{oquab2024dinov} compares two images in the CLIP/DINO model feature space, so they can tell whether two images are semantically similar and are widely used to evaluate the performance of personalized learning \citep{ruiz2023dreambooth,kumari2023multi,han2023svdiff}. Therefore, one intuitive solution for identifying unauthorized data usage in personalized diffusion models is to calculate the DINO score between the mimicries and the user's dataset, and claim infringement if the DINO score exceeds a certain threshold. However, we argue that CLIP/DINO score is inherently unreliable in our scenario, \textit{because a high feature-level similarity does not necessarily indicate piracy or the involvement of unauthorized data usage}. It could also be benign images with similar styles (\eg in the same art genre), or from independent models trained on similar (but authorized) data. For example, we trained a personalized model using the artwork of van Gogh and generated 40 mimicry images. We also collected 40 artworks from Raphael, an artist with a similar drawing style to van Gogh. We found that the DINO score cannot reliably distinguish between these two sets of images. It sometimes allocate a higher score to Raphael's artwork than the piracy model's mimicries, resulting in a very poor performance even with carefully-chosen thresholds. Therefore, we argue that DINO scores cannot be reliably applied to identify infringement in our setting. In contrast, \sys avoids this inherent limitation. This is because our identification is not drawn from the perspective of style similarity, but the existence of the unique external coating injected by the defender. Such a unique identifier is highly impossible to be coincidentally replicated in benign images, making our scheme more reliable. 

\vspace{0.3em}
\noindent\textbf{Discussion on the hypersphere classification.} As discussed in Section \ref{sec:hypersphere-classification}, raw binary classification may be suboptimal in our setting. This is because the positive training dataset is representative while the negative dataset is not. Specifically, all test-time samples are really positive (true positives) and follow a very similar distribution to the positive training set (\eg coated Pokemons), so the detector is expected to precisely identify those True Positives. However, real-world samples that should be considered as negative include not only uncoated Pokemon images, but also general artwork, and even out-of-distribution samples. However, our training set is not possible to cover all negative samples. For such out-of-distribution images, the behavior of the DNN becomes unpredictable \citep{Nguyen_2015_CVPR}. As such, the model may misclassify some actually negative (but domain-shifted) samples as positive (\ie false positives). In contrast, our method learns a hypersphere boundary that excludes all samples other than identified true positives as negative, it could generalize better on unseen negative samples, as evidenced in Table \ref{tab:classification-results}. 

\vspace{0.3em}
\noindent\textbf{Discussion on other possible solutions.} Besides DINO scores, there are also other methods possible to be applied to our problem. For example, passive methods such as membership/dataset/property inference can possibly determine data usage. However, these methods usually require white-box access to model weights or intermediate outputs, and may have both low TPR and high FPR \citep{wang2023diagnosis}. We believe this is due to the inherent complexity and generalizability of large-scale diffusion models. For proactive methods, we note that the watermark-based method in \citep{yu2021artificial} was originally designed for DeepFake attribution, but it has been adapted to the data usage verification problem without any adjustment since it only requires modifying the training data \citep{wang2023diagnosis}. Other image watermarks may also apply to our problem, but we believe they may suffer from similar issues. Notably, \citep{zhao2023recipe} uses the same watermark encoder/decoder with \citep{yu2021artificial}, so its performance will be the same as to \citep{yu2021artificial}. Therefore, we argue existing works are not sufficient for this challenging task.

\begin{figure*}[t]
    \centering
    \includegraphics[width=\linewidth]{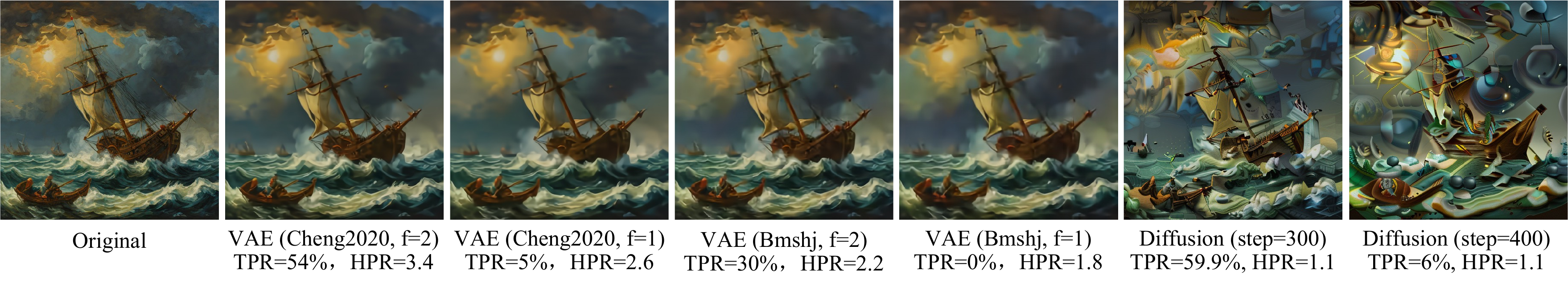}
    \vspace{-1.5em}
    \caption{\small Purification-based attack results. $\alpha$ is set to $10^{-9}$, and HPR refers to the human preference rating. Successful purification-based attacks introduce unnatural blurry areas, loss of details, or significant semantic changes that severely harm mimicry quality. Almost all images lose details of the mast. Please zoom in to better view the details.}
    \label{fig:purification-results}
\end{figure*}

\vspace{0.3em}
\noindent\textbf{Discussion on watermark removal attacks.} Current literature has proposed attacks against image watermarks \citep{JiangZG23,xu2023invisible}, which we broadly classify into box-free attacks (attacks that do not rely on any assumptions about the watermarking scheme, such as \citep{xu2023invisible}), and white/black box attacks (attacks that rely on access to watermark decoder, either white box or black box, including transfer-based attacks, such as \citep{JiangZG23}). As our method is not strictly a multi-bit watermark, we primarily focused on \citep{xu2023invisible} in our main paper. We also evaluated the transfer-based WEvade-B-S attack of \citep{JiangZG23} against {\sc Siren}. We find it is not effective against our method (under mild constraints). We believe this is possibly because the extractor is dependent on the training dataset, so it may exhibit limited transferability. We leave more explorations future work.

\vspace{0.3em}
\noindent\textbf{Discussion on trigger prompts for coating removal.} A possible attack against our method is to introduce a trigger prompt, \eg ``watermark'' into the training prompt of coated images, and remove such trigger prompt during generation. This may lead the diffusion model to attribute the coating features to the trigger prompt and decouple {\sc Siren}'s coating from the image. To verify whether this method defeats {\sc Siren}, we randomly select 400 images from Pokemon and coat them, and append ``with watermark" onto the training prompt. We assume the attacker has the other 433 uncoated Pokemon images with the original prompt. Then we personalize a model using these images. We find \sys is still successful in identifying infringement in the mimicries (generated using ``an image of a Pokemon"), with a TPR of 100. We hypothesize the failure of this attack is because the coating features are optimized to be regarded as the Pokemon's feature by the model with the learnability loss, rather than ``watermark". Therefore, even with the prompt ``watermark", the model still tends to regard it as the Pokemon's feature and fails to exclude it during learning. 

\vspace{0.2em}
\noindent\textbf{How to determine $\alpha$ in the real-world?} An interesting question is how to determine $\alpha$ in the real-world. As we know, larger $\alpha$ misses less true infringements but may also increase false claims, which is a trade-off. According to statistics \citep{Hampton2020}, there are $\sim$5,000 cases that are related to copyright infringement annually. Therefore, we humbly believe $\alpha=10^{-4}$ is generally sufficient and users may set $\alpha=10^{-9}$  for higher reliability. Nevertheless, we believe the choice of $\alpha$ should ideally be guided by future legal regulations and presiding judges in individual cases.

\vspace{0.5em}
\noindent\textbf{Discussion on BLIP-generated prompts.} One concern on our evaluation is the use of BLIP-generated prompts, as recent literature \citep{alemohammad2024selfconsuming} has shown that AI-generated data may lead to mode collapse and degrade training performance. However, we would like to clarify that the main finding of \citep{alemohammad2024selfconsuming} is that if we use purely AI-generated data to recursively train generative models, it would lead to model collapse. The key reason is that this process iteratively diminishes data quality and diversity. However, in our evaluation, BLIP is only used to generate image captions. The image data is real (instead of purely AI-generated), and the data is only used once (instead of recursively used). Therefore, the data diversity and quality are kept, and thus the aforementioned collapse problem would not happen. The BLIP-generated captions are shown to be highly effective in training diffusion models and vision language models \citep{li2023blip}, suggesting its effectiveness. It would be interesting to further discover the mechanisms and boundaries of AI-generated data's effectiveness, which we leave for future work.

\vspace{0.5em}
\noindent\textbf{Discussion on the characteristics of {\sc Siren}'s coating.} We are also interested in exploring the characteristics of {\sc Siren}'s coating. Note that $\mathcal{L}_\text{DM}$ is not updated during optimization so it would not introduce bad features that harm images' features. Instead, as the coating is optimized to be perceptually small and imperceptible, it does not negatively impact the image's style observed by HVS, so the generation quality is well preserved. We observe that the coating indeed encourages the alignment between text and image, indicated by a slightly higher text-image similarity measured by the CLIP model. This suggests \sys indeed introduces some features that can enhance the subject of the original image. As there are only limited tools for explaining diffusion models currently, we believe it would be interesting to further explore the underlying characteristics of \sys through both explainable AI tools and theoretical proofs in the future.

\newpage 

\section{Meta-Review}

The following meta-review was prepared by the program committee for the 2025
IEEE Symposium on Security and Privacy (S\&P) as part of the review process as
detailed in the call for papers.

\subsection{Summary}
This paper addresses unauthorized data use in personalized text-to-image diffusion models by introducing SIREN. It embeds imperceptible coatings into datasets to enable traceability in generated images, and these coatings are optimized to be relevant to the personalization task while maintaining image quality.

\subsection{Scientific Contributions}
\begin{itemize}
\item Provides a Valuable Step Forward in an Established Field
\end{itemize}

\subsection{Reasons for Acceptance}
\begin{enumerate}
\item The paper provides a valuable step forward in an established field. It addresses a relevant and significant problem of protecting personal data in the context of personalized diffusion models, by extending the data usage traceability to personalized model domain.
\item The paper provides an insight to explain the poor performance of existing methods (i.e. watermark-based methods and backdoor-based methods): in fine-tuning and personalized learning settings, the model is already primed to recognize meaningful signals from the images, causing it to ignore random coatings. This inspires the novel methodology of optimizing the coating as a feature recognizable by the model.
\end{enumerate}




\end{document}